\documentclass[aps,preprint,prd,nofootinbib]{revtex4-1}
\pdfoutput=1
\usepackage{graphicx}
\usepackage{psfrag}
\usepackage{float}
\usepackage{subfigure}
\usepackage{array}
\usepackage{graphicx}
\usepackage{amsmath}
\usepackage{amssymb}
\usepackage{mathrsfs}
\usepackage{bm}
\usepackage{slashed}
\usepackage{threeparttable}
\usepackage{multirow}
\usepackage[colorlinks, linkcolor=black, anchorcolor=black, citecolor=black]{hyperref}
\usepackage[amsmath,thmmarks,hyperref]{ntheorem}
\usepackage{soul}
\allowdisplaybreaks[4]

\begin{document}

\title{The study of light invisible particles in $B_c$ decays}
\author{Geng Li\footnote{karlisle@hit.edu.cn}, Tianhong Wang\footnote{thwang@hit.edu.cn}, Yue Jiang\footnote{jiangure@hit.edu.cn}, Xiao-Ze Tan\footnote{xz.tan@hit.edu.cn} and Guo-Li Wang\footnote{gl\_wang@hit.edu.cn}\\}
\address{Department of Physics, Harbin Institute of Technology, Harbin, 150001, China}

\baselineskip=20pt

\begin{abstract}

In this paper, we study the light scalar and pseudoscalar invisible particles in the flavor changing neutral current processes of the $B_c$ meson. Effective operators are introduced to describe the couplings between quarks and light invisible particles. The Wilson coefficients are extracted from the experimental results of the $B$ and $D$ mesons, which are used to predict the upper limits of the branching fractions of the similar decay processes for the $B_c$ meson. The hadronic transition matrix element is calculated with the instantaneously approximated Bethe-Salpeter method. The upper limits of the branching fractions when $m_\chi$ taking different values are presented. It is found that at some region of $m_\chi$, the channel $B_c\to D_s^{(\ast)}\chi\chi$ has the largest upper limit which is of the order of $10^{-6}$, and for $B_c\to D_s^\ast\chi\chi^\dagger$, the largest value of the upper limits can achieve the order of $10^{-5}$. Other decay modes, such as $B_c\to D^{(*)}\chi\chi^{(\dagger)}$ and $B_c\to B^{(*)}\chi\chi^{(\dagger)}$, are also considered.

\end{abstract}

\maketitle

\section{Introduction}

The Standard Model (SM) is extremely successful. However, it is considered to be an effective field theory which is valid only up to certain energy scale. For example, it will be invalid at the Planck scale, with gravity giving large contribution. Far below that, there are many arguments supporting that new physics (NP) will appear at the TeV scale. The NP can show itself as the missing energy in the collision at the $pp$ or $e^+e^-$ colliders. For example, CODEX-b at the LHCb experiment is proposed to probe for GeV-scale long-lived particles~\cite{Gligorov:2017nwh}. If we assume the possible new particle to be the candidate for the dark matter (DM), the high energy collision will provide a powerful way to detect such particles. Among the DM candidates, the weakly interacting massive particle (WIMP), which appears in many theoretical models, has attracted extensive attention (see \cite{Roszkowski:2017nbc} for reviews). The WIMP annihilation cross section is constrained by the observed dark matter density, which sets the lower bound of the WIMP mass to a few GeV (the so-called Lee-Winberg limit~\cite{PhysRevLett.39.165}). However, this result is model-dependent. If the DM is nonfermionic and the weak mass scales or weak interactions are not assumed~\cite{Feng:2008ya}, this constraint can be relaxed, and more lower mass, such as a few keV, will be possible. Theoretically, this kind of light dark matter (LDM) can have different spins\cite{Belyaev:2018pqr}, for example, it can be a scalar particle~\cite{Boehm:2003hm}, sterile neutrino~\cite{Kusenko:2009up}, or hidden vector particle~\cite{Hambye:2008bq}. The MeV-scale LDM is proposed~\cite{Pospelov:2007mp,Hooper:2008im} to explain the unexpected emission of 511 keV photons from the galaxy center. Experimentally, the parameter space for the WIMP with mass larger than several GeV has been severely constrained by the recent experiment~\cite{Aprile:2017iyp}, which also provides a strong motivation for the study of the sub-GeV dark matter.  

The LDM emission from the heavy meson decays is an interesting approach for such studies. Phenomenologically, the LDM of some hidden sector can weakly interact with the SM fermions through different ways. For example, it can couple directly to the Higgs boson~\cite{Kim:2009qc,Winkler:2018qyg}. Or there are some connectors with quantum numbers of both SM and hidden sectors. Such connector can be a chiral fermion~\cite{McKeen:2009rm} or a dark gauge boson~\cite{Darme:2017glc}. At the energy level of heavy mesons, these processes will be greatly suppressed by the large mass in the propagator of the connector or by the small coupling constant between the connector and the SM fermions. By a model-independent way, we can introduce an effective Lagrangian to describe phenomenologically the interaction between the invisible particles and SM fermions. This method has been extensively used in Refs.~\cite{Bird:2004ts,Bird:2006jd,Badin:2010uh,Gninenko:2015mea,Barducci:2018rlx,Kamenik:2011vy,Bertuzzo:2017lwt} to study the flavor-changing neutral current (FCNC) processes of $K$, $D$, and $B$ mesons. The SM background comes from the decays with $\nu\bar\nu$ in the final states, which has small branching fraction and makes the detection of NP possible. The difference between the experimental results for $M\to M_f\slashed E$ and the theoretical predictions for $M\to M_f\nu\bar\nu$ in the SM will set the constrains for the LDM emission channels, where $M$ and $M_f$ are the masses of the initial and final mesons, respectively. 

The same analysis can also be applied to the $B_c$ meson. As consisting of a heavy quark and a heavy antiquark with different flavors, this meson is unique. It can only decay through weak interaction, and either the $b$ quark or the $\bar c$ antiquark can be a spectator. Therefore more possible decay modes involving the invisible particles are allowed. Experimentally, there are abundant $B_c$ samples are collected at the LHC~\cite{Aaij:2014ija,Gouz:2002kk,Ivanov:2005fd}, which gives us the chance to study its various decay channels precisely, especially the rare decays. Until now there is no experimental data for such decays of the $B_c$ meson available, so we expect detections in the near future. Theoretically, many methods have been applied to study the semileptonic, nonleptonic, pure leptonic, and the FCNC processes of the $B_c$ meson~\cite{Kiselev:2000pp,Kiselev:2001zb,Choi:2009ym,Ebert:2010zu,Rui:2011qc}. In this work, we will apply the instantaneous Bethe-Salpeter method to calculate the hadronic transition amplitude when both the initial and final mesons are heavy. This method has been used extensively to study the weak decays of $B_q$ mesons~\cite{Zhang:2010ur,Fu:2011zzo}, and gotten consistent results with experiments. In the LDM emission processes of the $B_c$ meson, this method is still valid and the calculation steps are the similar to those in the SM.

The rest of the paper is organized as follows: In Sec. II, we first construct the effective Lagrangian which describes the coupling between quarks and light invisible particles. Then by comparing the theoretical and experimental results, we extract the upper and lower limits of the Wilson coefficients. In Set. III,  these limits are used to constrain the branching fractions of the decay channel $B_c\to h\chi\chi$ with $h$ and $\chi$ being the final meson and the invisible particle, respectively. Finally, we give the summary and perspective in Sec. IV. 

\section{Effective operators }

\subsection{$\chi$ is a scalar}
\begin{figure}[htb]
	\centering
	\subfigure[]{\label{}
		\includegraphics[width=0.42\textwidth]{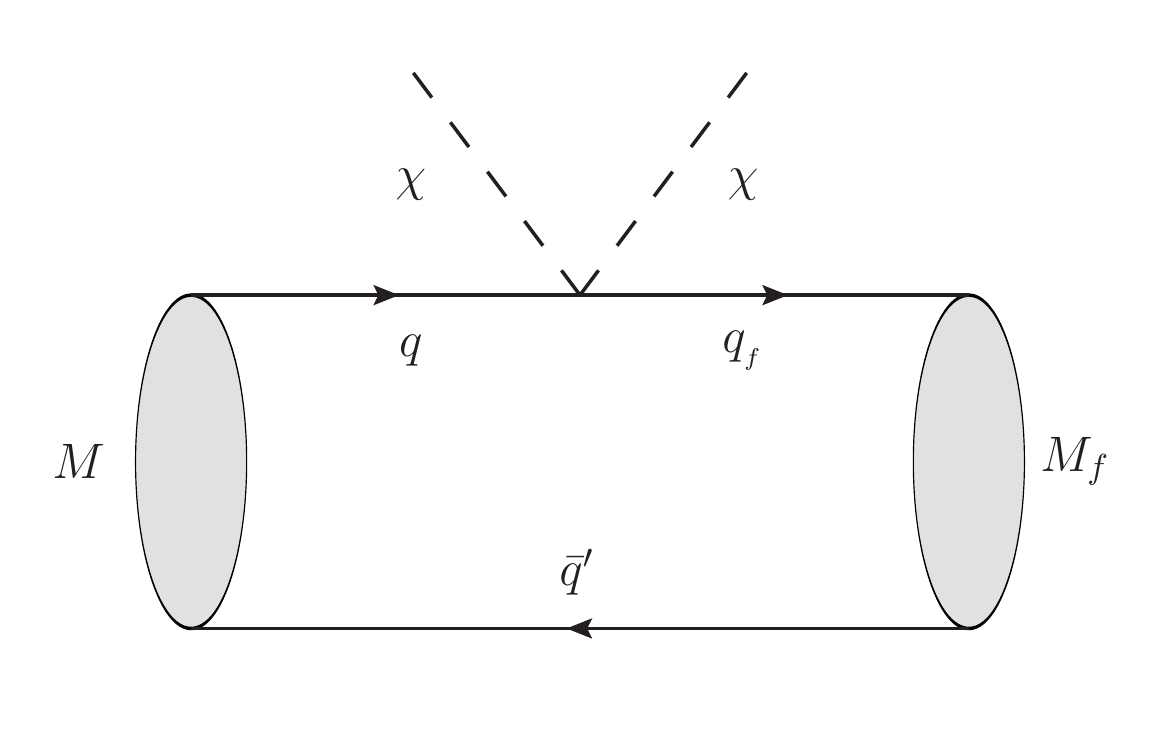}}
	\subfigure[]{\label{} 
		\includegraphics[width=0.33\textwidth]{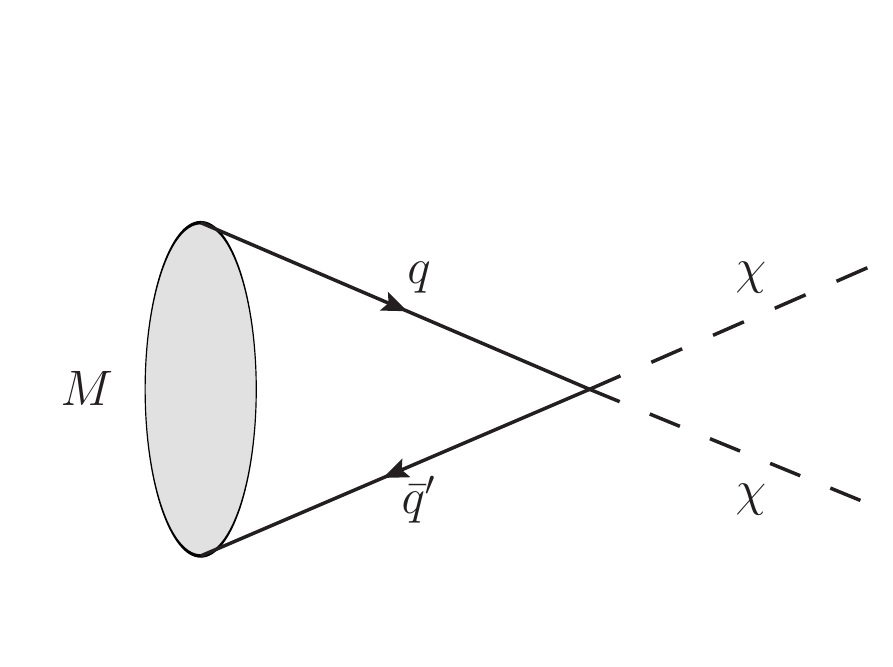}} \\
	\caption{Feynman diagrams of decay channels involving invisible particles.}
	\label{fig1}
\end{figure}

At the quark level, the $\chi$ emission processes of the heavy meson can be described by the effective Lagrangian~\cite{Badin:2010uh},
\begin{equation}
\begin{aligned}
\mathcal L_{1}&=g_{s1} m_q(\bar q_{_f} q)(\chi\chi)+g_{s2} m_q(\bar q_{_f}\gamma^{5} q)(\chi\chi),
\label{eq1}
\end{aligned}
\end{equation}
where $q$ and $q_{_f}$ are the Dirac spinor fields of the initial and final quarks, respectively; $g_{s1}$ and $g_{s2}$ are the phenomenological coupling constants. This Lagrangian is model-independent. And for specific models, the four-particle vertex may be generated at the tree or loop level~\cite{Bird:2004ts,Bird:2006jd,Badin:2010uh,Gninenko:2015mea,Barducci:2018rlx} by introducing other new particles.  In this work, we will not focus on any specific model, but consider the FCNC processes of $B_c$ meson induced by such effective operators. Theoretically, there are many studies~\cite{Ebert:2010dv, Choi:2010ha, Geng:2001vy, Wang:2014yia} of the FCNC processes of the $B_c$ meson, while the corresponding detection is still missing. So we cannot use the experimental data of $B_c$ meson to set constraints on the coupling constants. Our strategy is in the opposite direction. That is, the allowed-region of the coupling constants from other processes are used to constraint the branching ratios of the $B_c$ decays. Experimentally, there are data for such decays of $B$ and $D$ mesons. The corresponding channels are $ B ^ {-} \to K ^ -(K^{\ast-}) + \slashed E $, $ B ^ {-} \to \pi ^ -(\rho^-) + \slashed E $, and $ B ^ {0} \to \slashed E $ for $B$ meson, and $ D ^ {0} \to \slashed E $ for $D$ meson. The experimental bounds for their branching ratios are listed in Table \ref{tab1}.  Within the SM, the missing energy $\slashed E$ represents the $\nu\bar\nu$ pair, and the branching fractions are calculated in Refs.~\cite{Kamenik:2009kc,Jeon:2006nq,Altmannshofer:2009ma,Bartsch:2009qp}. The difference between theoretical predictions and experimental bound allows the existence of NP. Here the NP processes are described by the Feynman diagrams in Fig. \ref{fig1}. 
\begin{table}[htb]
	\setlength{\tabcolsep}{0.5cm}
	\caption{The branching ratios (in units of $10^{-6}$) of $B$ and $D$ decays involving missing energy.}
	\centering
	\begin{tabular*}{\textwidth}{@{}@{\extracolsep{\fill}}ccc}
		\hline\hline
		Experimental bound~\cite{Chen:2007zk,Grygier:2017tzo,Lai:2016uvj}&SM prediction~\cite{Kamenik:2009kc,Jeon:2006nq,Altmannshofer:2009ma,Bartsch:2009qp}&Invisible particles bound\\
		\hline
		${\rm BR}(B^\pm\to K^\pm\slashed E)<14$& ${\rm BR}(B^\pm\to K^\pm\nu \bar{\nu})=5.1 \pm 0.8$  & ${\rm BR}(B^\pm\to K^\pm\chi\chi)<9.7$	\\
		${\rm BR}(B^\pm\to \pi^\pm\slashed E)<14$& ${\rm BR}(B^\pm\to \pi^\pm\nu \bar{\nu})=9.7 \pm 2.1$  & ${\rm BR}(B^\pm\to \pi^\pm\chi\chi)<6.4$	\\
		${\rm BR}(B^\pm\to K^{*\pm}\slashed E)<61$& ${\rm BR}(B^\pm\to K^{*\pm}\nu \bar{\nu})=8.4 \pm 1.4$  & ${\rm BR}(B^\pm\to K^{*\pm}\chi\chi)<54$	\\
		${\rm BR}(B^\pm\to \rho^\pm\slashed E)<30$&${\rm BR}(B^\pm\to \rho^\pm \nu \bar{\nu})=0.49^{+0.61}_{-0.38}$  & ${\rm BR}(B^\pm\to \rho^\pm \chi\chi)<30$	\\
		${\rm BR}(B^0\to \slashed E)<47$&${\rm BR}(B^0\to \nu \bar{\nu})\sim 0$  & ${\rm BR}(B^0\to \chi\chi)<47$	\\
		${\rm BR}(D^0\to \slashed E)<94$&${\rm BR}(D^0\to \nu \bar{\nu})\sim 0$  & ${\rm BR}(D^0\to \chi\chi)<94$	\\
		\hline\hline
	\label{tab1}
	\end{tabular*}
\end{table}

For the $ B^ - \to \pi^ - (K^-)\chi\chi$ processes, only the scalar current gives contribution to the transition amplitude, which can be written as
\begin{equation}
\begin{aligned}
\langle K^-(\pi^{-})\chi\chi|\mathcal L_{1}|B^-\rangle&=2 g_{s1} m_q\frac{(P-P_f)_\mu}{m_q-m_{q_{_f}}} \langle K^-(\pi^{-})|(q_{_f}\gamma^\mu q) |B^-\rangle\\
&= 2 g_{s1} m_q\frac{(P-P_f)_\mu}{m_q-m_{q_{_f}}}\bigg \{ (P+P_f)^{\mu}f_+ (s)+(P-P_f)^{\mu} \frac{M^2-M_{f}^2}{s}\big[f_0 (s)-f_+(s)\big]\bigg\}\\
&= \frac{2 g_{s1} m_q}{m_q-m_{q_{_f}}}(M^2-M_{f}^2)f_0 (s),
\label{eq2}
\end{aligned}
\end{equation}
where $P$ and $P_f$ are the momenta of the initial or final mesons, respectively; $m_q$ and $m_{q_f}$ are the masses of quarks;  $s$ is defined as $(P-P_f^{})^2$. In the first step, the equation of motion is used. The hadronic transition matrix is parameterized as the form factors $f_+$ and $f_0$. Here we adopt the results of the QCD light-cone sum rules (LCSR)~\cite{Ball:2004ye}, where the form factors are constructed as 
\begin{equation}
\begin{aligned}
&f_0(s)=\frac{r_2}{1-s/m_{fit}^2},\\
&f_+^K(s)=\frac{r_1}{1-s/m_R^2}+\frac{r_2}{(1-s/m_R^2)^2},\\
&f_+^{\pi}(s)=\frac{r_1}{1-s/m_R^2}+\frac{r_2}{1-s/m_{fit}^2}.\\
\label{eq3}
\end{aligned}
\end{equation}
The corresponding parameters are presented in Table \ref{tab2}.
\begin{table}[htb]
	\setlength{\tabcolsep}{0.5cm}
	\caption{Parameters in the form factors of the $B\to \pi(K)$ processes~\cite{Ball:2004ye}.}
	\centering
	\begin{tabular*}{\textwidth}{@{}@{\extracolsep{\fill}}ccccc}
		\hline\hline
		$F_i$&$r_1$&$r_2$&$m_{fit}^2$ (GeV$^2$)&$m_R$ (GeV)	\\
		\hline
		$f^K_0$&$0$&$0.330$&$37.46$&$-$         \\
		$f^K_+$&$0.162$&$0.173$&$-$&$5.41$    \\
		\hline
		$f^{\pi}_0$&$0$&$0.258$&$33.81$   &$-$    \\
		$f^{\pi}_+$&$0.744$&$-0.486$&$40.73$&$5.32$  \\
		\hline\hline
	\label{tab2}
	\end{tabular*}
\end{table}

For the $ B^ - \to \rho^ - (K^{\ast-})\chi\chi$ processes, only the pseudoscalar current gives contribution to the transition amplitude, which has the form,
\begin{equation}
\begin{aligned}
\langle K^{*-}(\rho^{-})\chi\chi|\mathcal L_{1}|B^-\rangle&=2 g_{s2} m_q\frac{(P-P_f)_\mu}{m_q-m_{q_{_f}}} \langle K^*(\rho)^{-}|(q_{_f}\gamma^\mu \gamma^5 q) |B^-\rangle\\
&= i\frac{2 g_{s2}m_q }{m_q-m_{q_f^{}}} (P-P_f)_{\mu}\bigg\{\epsilon^{\mu} (M+M_f)A_1(s)-(P+P_f)^{\mu}\big[\epsilon \cdot (P-P_f)\big]\\
&~~~\times\frac{A_2(s)}{(M+M_f)}-(P-P_f)^{\mu}\big[\epsilon \cdot (P-P_f)\big]\frac{2M_f}{(P-P_f)^2}\big[A_3(s)-A_0(s)\big]\bigg\}\\
&= i\frac{4 g_{s2}m_q M_f}{m_q-m_{q_f^{}}} \big[\epsilon \cdot (P-P_f)\big]  A_0(s),
\label{eq4}
\end{aligned}
\end{equation}
where $\epsilon$ is the polarization vector of the final meson; $M$ and $M_f$ are the masses of the initial and final mesons, respectively; $A_0$, $A_1$, $A_2$, and $A_3$ are form factors.

In Ref.~\cite{Straub:2015ica}, Bharucha {\sl et al.} also used the LCSR, but with a different parameterization method, to write the form factors as 
\begin{equation}
F_i(s) = P_i(s) \sum_k \alpha_k^i \,\left[z(s)-z(0)\right]^k,
\label{eq5}
\end{equation}
where $F_1,~F_2,~F_3$, and $F_4$ represent $A_0,~A_1,~A_{12}$, and $V$, respectively; $P_i(s)=(1-s/m_{R,i}^2)^{-1}$ represents the pole structure. And $z(s)$ is defined as 
\begin{equation}
z(s) = \frac{\sqrt{s_+-s}-\sqrt{s_+-s_0}}{\sqrt{s_+-s}+\sqrt{s_+-s_0}},
\label{eq6}
\end{equation}
where
$s_\pm \equiv (M\pm M_f)^2$ and $s_0\equiv s_+(1-\sqrt{1-s_-/s_+})$. The related parameters are listed in Table \ref{tab3}.
The form factors $A_{12}$ and $A_3$ are related to $A_1$ and $A_2$ by 
\begin{equation}
\begin{aligned}
A_{12}(s)&=  \frac{ (M+M_f){}^2 (M^2-M_f^2-s)A_1(s) - \big [(M+M_f)^2-s\big] \big[ (M-M_f)^2-s \big] A_2(s)  }{16 M M_f^2 (M+M_f)},\\
A_3(s)&=\frac{M+M_f}{2M_f}A_1(s)-\frac{M-M_f}{2M_f}A_2(s).
\label{eq7}
\end{aligned}
\end{equation}
\begin{table}[htb]
	\setlength{\tabcolsep}{0.5cm}
	\caption{Parameters in the form factors of the $B\to \rho(K^*)$ processes with $k_{\rm max}=2$~\cite{Straub:2015ica}.}
	\label{}
	\centering
	\begin{tabular*}{\textwidth}{@{}@{\extracolsep{\fill}}ccccc}
		\hline\hline
		$F_i$& $B\to K^*$ &$m_{R,i}^{b \to d}/$GeV& $B\to\rho$&$m_{R,i}^{b \to s}/$GeV \\
		\hline
		$\alpha_0^{A_0}$ & $0.36 \pm 0.05$ && $0.36 \pm 0.04$ &\\
		$\alpha_1^{A_0}$ & $-1.04 \pm 0.27$ &$5.279$& $-0.83 \pm 0.20$&$5.366$  \\
		$\alpha_2^{A_0}$ & $1.12 \pm 1.35$ && $1.33 \pm 1.05$ &\\
		\hline
		$\alpha_0^{A_1}$ & $0.27 \pm 0.03$ && $0.26 \pm 0.03$ &\\
		$\alpha_1^{A_1}$ & $0.30 \pm 0.19$ &$5.724$& $0.39 \pm 0.14$ &$5.829$\\
		$\alpha_2^{A_1}$ & $-0.11 \pm 0.48$ && $0.16 \pm 0.41$ &\\
		\hline
		$\alpha_0^{A_{12}}$ & $0.26 \pm 0.03$ && $0.30 \pm 0.03$ &\\
		$\alpha_1^{A_{12}}$ & $0.60 \pm 0.20$ &$5.724$& $0.76 \pm 0.20$  &$5.829$\\
		$\alpha_2^{A_{12}}$ & $0.12 \pm 0.84$ && $0.46 \pm 0.76$ &\\
		\hline
		$\alpha_0^{V}$ & $0.34 \pm 0.04$ && $0.33 \pm 0.03$ &\\
		$\alpha_1^{V}$ & $-1.05 \pm 0.24$ &$5.325$& $-0.86 \pm 0.18$ &$5.415$\\
		$\alpha_2^{V}$ & $2.37 \pm 1.39$ && $1.80 \pm 0.97$ &\\
		\hline\hline
	\label{tab3}
	\end{tabular*}
\end{table}

By finishing the three-body phase space integral, we get the branching ratios
\begin{equation}
\mathcal {BR}  = \frac{1}{512  \pi^3 M^3 \Omega\Gamma_{B^-}}\int\frac {ds}{s} \lambda^{1/2}(M^2, s, M_f^2)\lambda^{1/2}(s, m_\chi^2, m_\chi^2)\int d\cos\theta\sum_\lambda|\mathcal M|^2,
\label{eq8}
\end{equation}
where $\lambda(x, y, z)= x^2 + y^2 +z^2 -2xy-2xz -2yz$ is the K${\rm \ddot a}$llen function; $m_\chi$ is the mass of the invisible particle; $\theta$ is the angel between the three-dimensional momenta $\vec P_\chi$ and $\vec P_f$ in the momentum center frame of invisible particles; $\Gamma_{B^-}$ is the total width of $B^-$ meson; $\Omega=2$ originates from the final two invisible particles being identical. 

\begin{figure}[htb]
	\centering
	\includegraphics[width=0.43\textwidth]{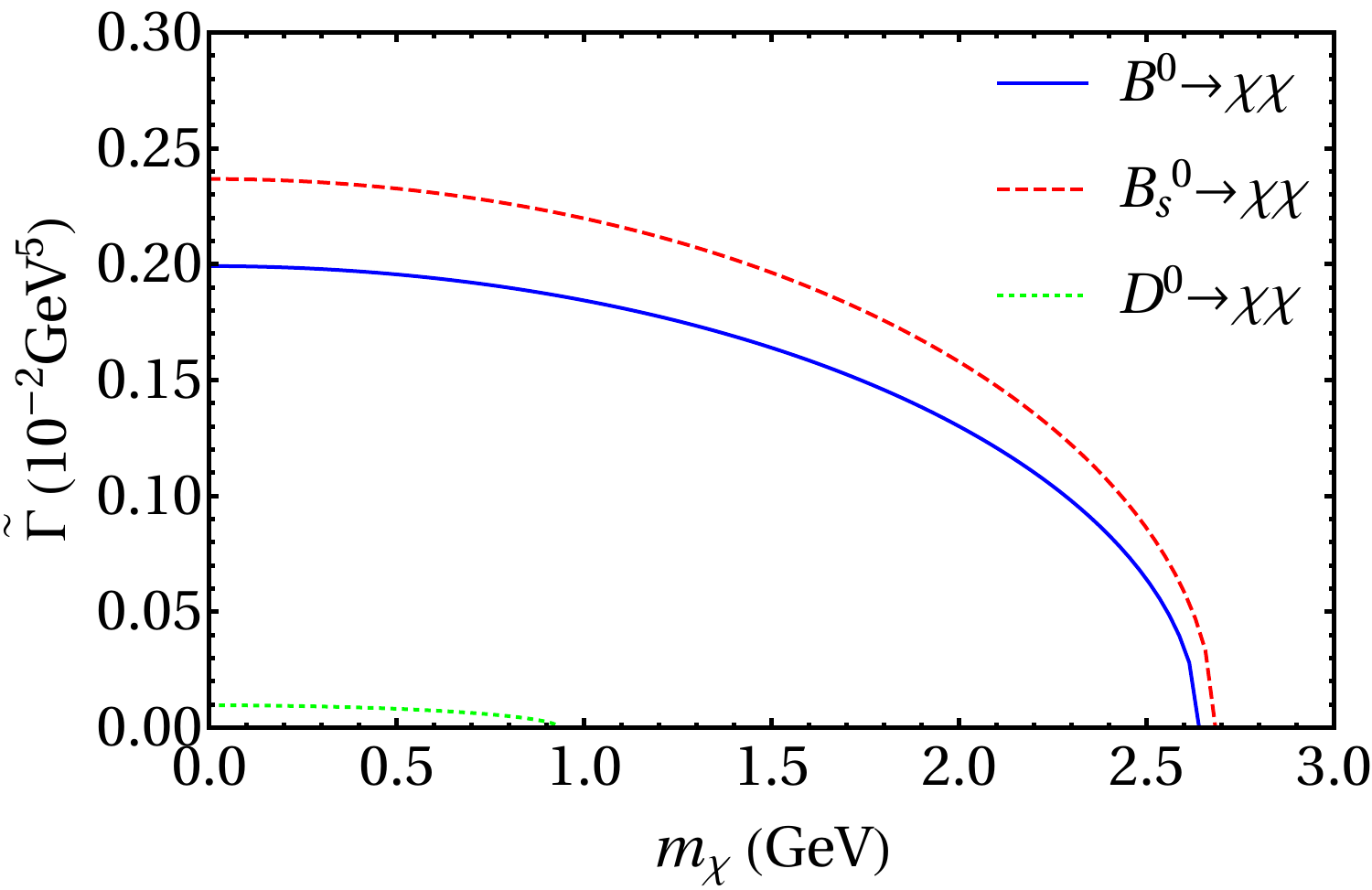}
	\caption{Meson annihilation modes.}
	\label{fig2}
\end{figure}

For the annihilation processes of $B^0$, $D^0$, $B_s^0$ mesons, that is $ M \to \chi \chi $, only the pseudoscalar current contributes to the decay amplitude, which has the form,
\begin{equation}
\begin{aligned}
\langle \chi\chi|\mathcal L_{1}|M\rangle 
=\frac{2g_{s2}m_q}{m_q+m_{\bar q}}M^2f_M,
\label{eq9}
\end{aligned}
\end{equation}
where $f_M$ is the decay constant of the initial meson, which has the values: $ f_ {B_u ^ 0} = 0.196 $ GeV, $ f_ {B^0_s} = 0.216 $ GeV and $f_ {D^ 0} = 0.230 $ GeV~\cite{Cvetic:2004qg}. By finishing the two-body phase space integral we get the partial width,
\begin{equation}
\begin{aligned}
\Gamma =\frac{1}{16\pi M \Omega}\sqrt{1-\frac{4m^2_{\chi}}{M^2}}|\mathcal M|^2.
\label{eq10}
\end{aligned}
\end{equation}
In Fig. \ref{fig2}, we plot $\tilde\Gamma=\Gamma/|g_{s2}|^2$ as a function of $m_\chi$. One can see that they all have same trend that decrease when $m_\chi$ gets larger, because the phase space gets smaller.

\begin{figure}[htb]
	\centering
	\subfigure[~$|g_{s1}|^2$]{
		\label{fig3a}
		\includegraphics[width=0.43\textwidth]{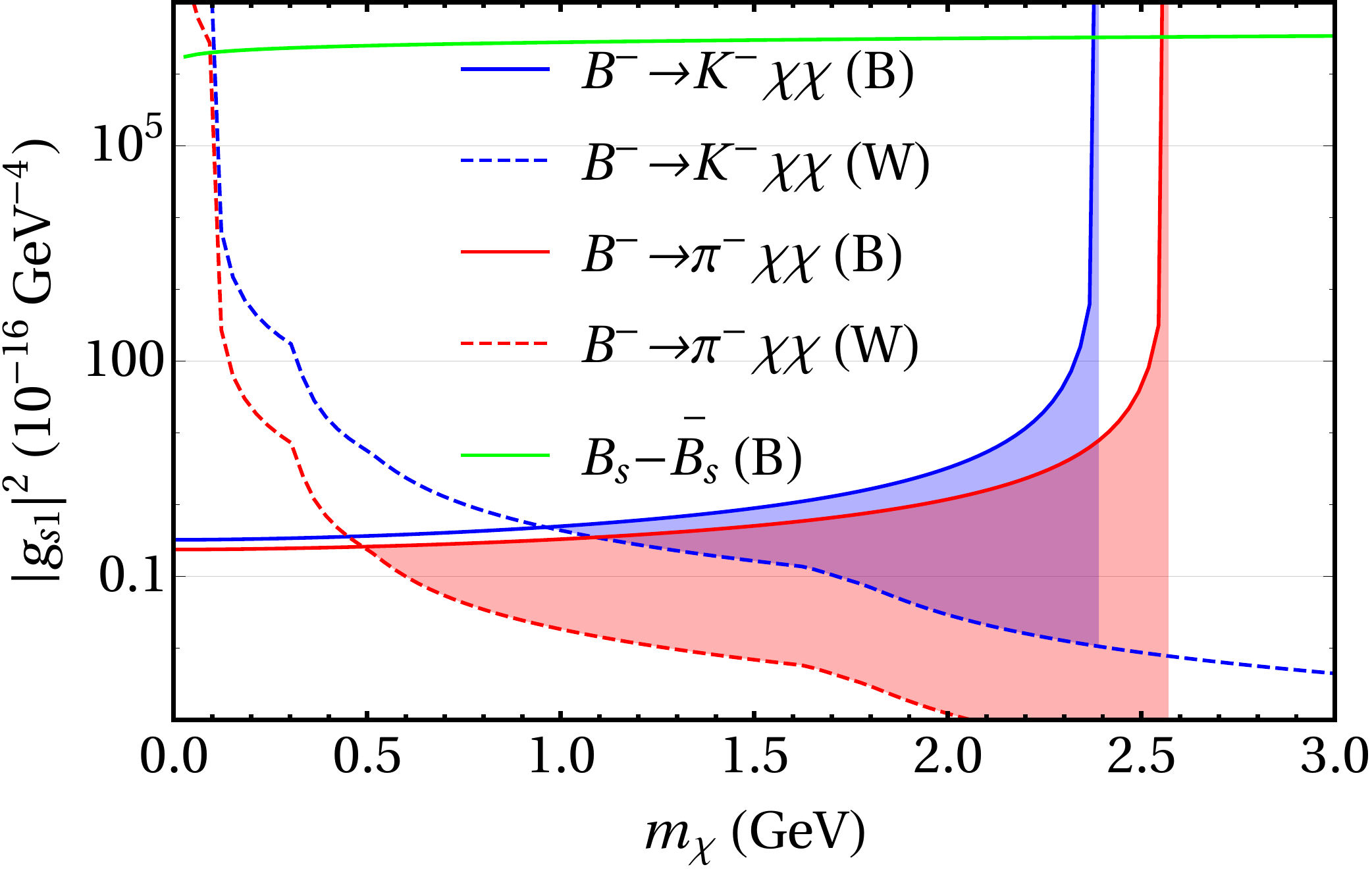}} 
		\hspace{0.5cm}
	\subfigure[~$|g_{s2}|^2$]{
		\label{fig3b}
		\includegraphics[width=0.43\textwidth]{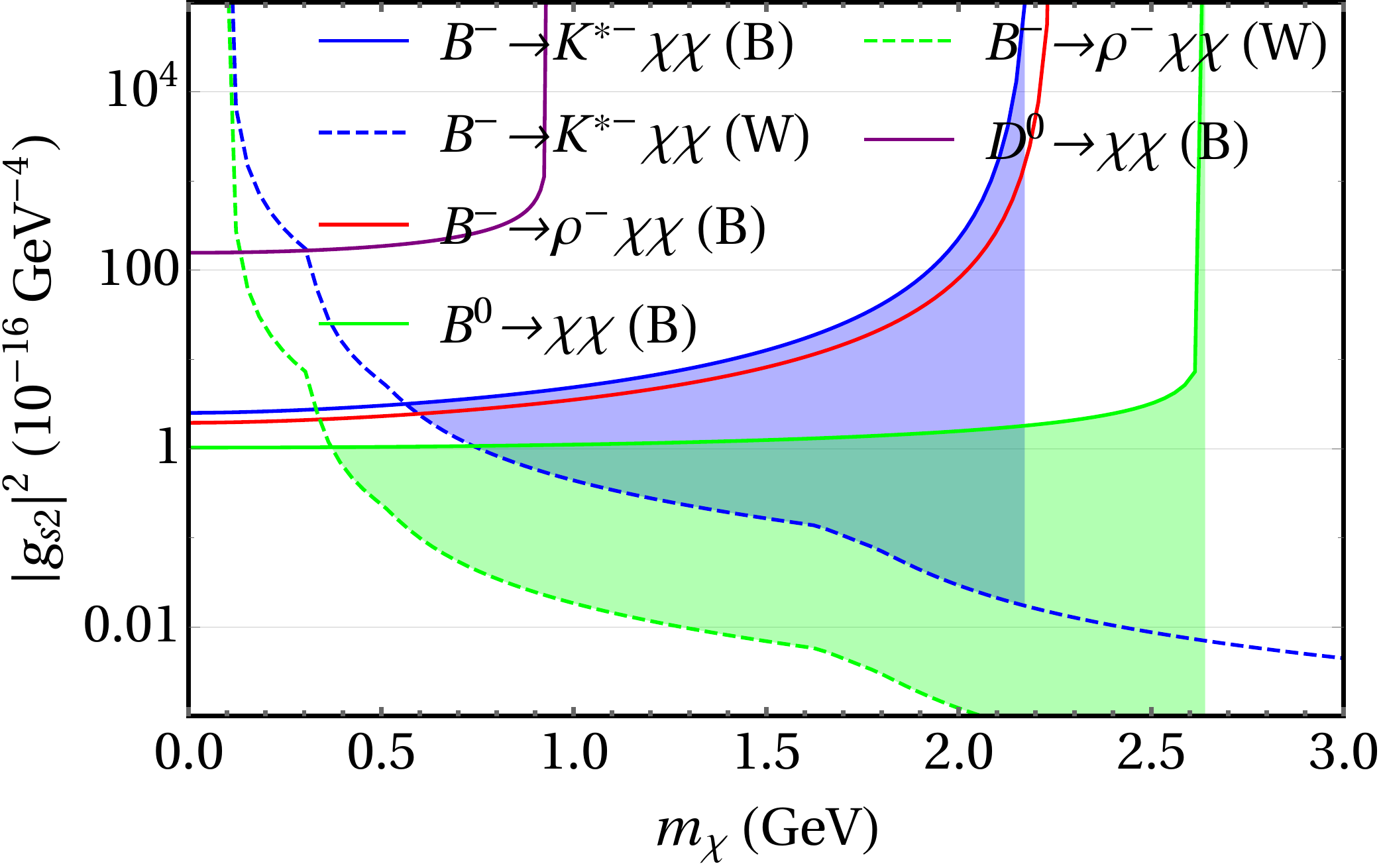}} \\
	\caption{The experimental bounds of $|g_{s1}|^2$ and $|g_{s2}|^2$ with different $m_\chi$. (B) represents Belle and (W) represents WMAP.}
	\label{fig3}
\end{figure}

By comparing the theoretical predictions and the experimental upper limits (the third column of Table \ref{tab1}) of the branching ratios for these decays, we can set the upper bounds for the effective coupling constants $g_{s1}$ and $g_{s2}$ with specific mass of the invisible particle. The results are shown in Fig. \ref{fig3} (represented by the solid lines).
One can see that as $m_\chi$ increasing, the upper limits of the effective coupling constants get more and more larger. The reason is simple: larger $m_\chi$ means more suppression from the phase space. So from these decay channels we can set more stringent upper limits for the effective coupling constants when $2m_\chi$ is not close to the threshold. The $B\to\pi(\rho)\chi\chi$ channel gives smaller bound of $|g_{s1}|^2$ ($|g_{s2}|^2$) compared with the $B\to K(K^\ast)\chi\chi$ channel. The $B\to\chi\chi$ mode gives the most stringent upper bound of $|g_{s2}|^2$, because the two-body phase space is larger than the three-body case. We also present the result from $D^0$ decay, which is larger due to its smaller mass. For $B_s^0$, the experimental results are still missing. Once the experimental data for the annihilation channel are available, they can also be used to set the upper limit of $|g_{s2}|^2$.

\begin{figure}[htb]
	\centering
	\includegraphics[width=0.4\textwidth]{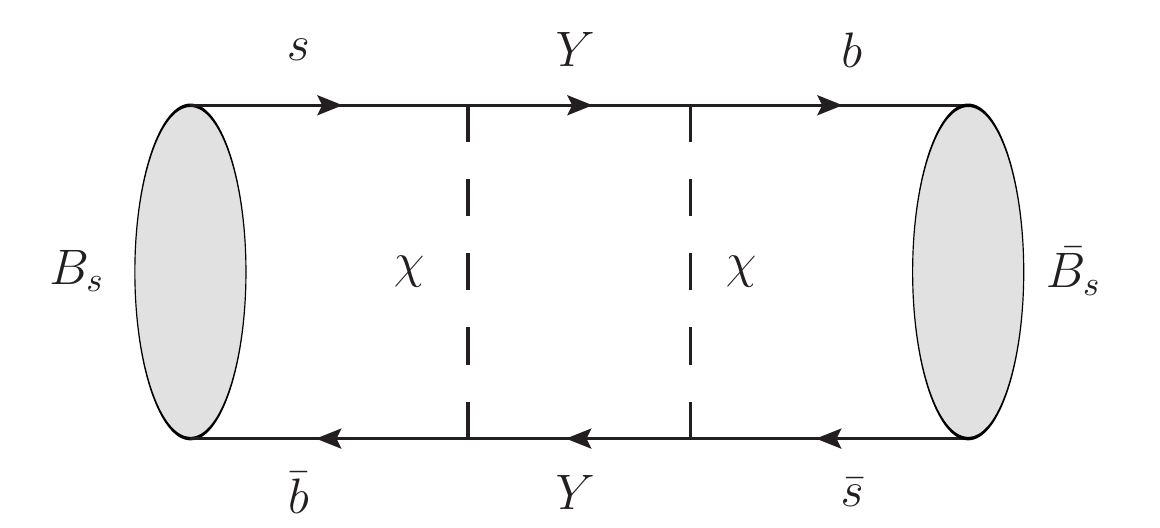}
	\caption{Feynman diagrams of meson mixing involving the invisible particles.}
	\label{fig4}
\end{figure}

The discussions above are model-independent except the calculations of the hadronic transition matrix. If we introduce some specific models, more information can be extracted. For example, in Ref.~\cite{McKeen:2009rm}, McKeen introduced an effective Lagrangian, 
\begin{equation}
\begin{aligned}
{\mathcal L_{\rm int}}= \Lambda_{f} \chi\bar Y_L f_R+\Lambda_f \chi\bar Y_R f_L,
\label{eq11}
\end{aligned}
\end{equation}
where $Y$ is a heavy fermion and $\Lambda$ is the coupling constant. With the $B_s^0-\bar{B_s^0}$ mixing (see Fig. \ref{fig4}), a mass difference can be estimated to be 
\begin{equation}
\begin{aligned}
\Delta M_{B_s-\bar{B_s}} \simeq \frac{|g_{s1}|^2m_q^2}{1152\pi^2}f_{B_s^0}^2M\bigg[8+5(\frac{M}{m_b+m_s})\bigg]\log(\frac{m^2_Y}{m^2_{\chi}}),
\label{eq12}
\end{aligned}
\end{equation}
where we have related $\frac{2\Lambda_b\Lambda_s^*}{m_Y}$ to $m_q g_{s1}$. Experimentally, the latest value of $\Delta M_{B_s-\bar{B_s}}$ is $17.63 \pm 0.11 ~{\rm (stat)} \pm 0.02 ~{\rm(syst)~ps}^{-1}$, which comes from LHCb~\cite{Aaij:2011qx}. This sets an upper limit for the contribution of the light invisible particle. And by assuming $m_Y\simeq400$ GeV~\cite{McKeen:2009rm} we can estimate the upper limit of $g_{s1}$ which is shown by the green solid line in Fig. \ref{fig3a}. One can see it's a very loose restriction compared with other decay modes. So we will not use this result in the $B_c$ case. 

The lower bound for the effective couplings can be set by considering the relic density~\cite{Bertone:2004pz, Komatsu:2008hk}
\begin{equation}
\Omega_c h^2 = 0.1131\pm 0.0034\ge \frac{0.1 {\rm pb}}{\langle \sigma_{\chi} v_{rel}\rangle},
\label{eq13}
\end{equation}
where $\chi$ is the candidate of dark matter which assumed to be a scalar and SM singlet. It can annihilate into the SM particles by introducing the effective Lagrangian~\cite{Bird:2004ts, Kim:2009qc}
\begin{equation}
\begin{aligned}
{\mathcal L}= \frac{1}{2}(\partial \chi)^2 -\frac{1}{2}m_{\chi}^2 {\chi}^2 -\frac{\Lambda_{\chi}}{4!}{\chi}^4 -\frac{\hat{\Lambda}}{2}{\chi}^2 H^\dagger H,
\label{eq14}
\end{aligned}
\end{equation}
where $H$ is the SM Higgs doublet.
As calculated in Ref.~\cite{Kim:2009qc}, the annihilation cross section has the form 
\begin{equation}
\begin{aligned}
\langle \sigma_{\chi} v_{rel}\rangle =\frac{\hat{\Lambda}^2 m_{\chi}^2}{ \pi m_h^4} \sum_f x_f^2 (1-x_f^2)^{3/2},
\label{eq15}
\end{aligned}
\end{equation}
where $x_f=m_f/m_{\chi}$ and $f$ refers to the SM fermions. Combining Eqs.~(13) and (15), we can get the lower limit for $\hat\Lambda$, which is related to $g_{s1}$ and $g_{s2}$ by~\cite{Kim:2009qc}
\begin{equation}
g_{s1}(g_{s2})= \frac{3G_Fm_t^2}{64\sqrt{2}\pi^2}\frac{\hat{\Lambda} V_{tb}^*V_{ts}}{m_h^2},
\label{eq16}
\end{equation}
 where $G_F$ is the Fermi coupling constant; $V_{q_1q_2}$ is the Cabibbo-Kobayashi-Maskawa (CKM) matrix element. So the lower limits of $g_{s1}$ and $g_{s2}$ are also obtained which are represented by the dashed lines in Fig. \ref{fig3}. The shadow areas is allowed by both constraints (meson decays and relic density). When the parameters are in this region, the invisible scalar particle can be a candidate of the DM. For the regions below the dashed lines, the scalar invisible particle can also be possible to exist as a portal DM, which is a mediator between SM and hidden sectors~\cite{Pospelov:2007mp,Patt:2006fw,Andreas:2010dz,Krnjaic:2015mbs}.

\subsection{$\chi$ is a pseudoscalar}

If $\chi$ is a pseudoscalar,  $\chi$ and $\chi^\dagger$ represent different fields. The effective Lagrangian which describes the FCNC processes $q\to q_f\chi\chi^\dagger$ has the form~\cite{Badin:2010uh},
\begin{equation}
\begin{aligned}
\mathcal L_{2}&=g_{p1} m_q(\bar q_{_f} q)(\chi^\dagger\chi)+g_{p2} m_q(\bar q_{_f}\gamma^{5} q)(\chi^\dagger\chi)+g_{p3} (\bar q_{_f}\gamma^\mu  q)(\chi^\dagger\overset{\leftrightarrow}{\partial}_\mu\chi)\\
&~~~+g_{p4} (\bar q_{_f}\gamma^\mu  \gamma^5 q)(\chi^\dagger\overset{\leftrightarrow}{\partial}_\mu\chi),
\label{eq17}
\end{aligned}
\end{equation}
where we have used the definition $\chi^\dagger\overset{\leftrightarrow}{\partial}_\mu\chi\equiv\chi^\dagger(\partial_\mu\chi)-(\partial_\mu\chi^\dagger)\chi$. The last two terms disappear when $\chi$ is a scalar.

For the decays of $B$ meson, when the final meson is a pseudoscalar, the second and the fourth terms in Eq. (\ref{eq17}) will not contribute to the decay. The FCNC process can be induced by the scalar or vector current, and the transition amplitude has the form
\begin{equation}
\begin{aligned}
\langle h^-\chi^\dagger\chi|\mathcal L_{2}|B^-\rangle=\Big[\frac{g_{p1} m_q}{m_q-m_{q_{_f}}}(P-P_f)_\mu + g_{p3} (P_1-P_2)_\mu\Big] \langle h^-|(\bar {q_{_f}}\gamma^\mu  q) |B^-\rangle,
\label{eq18}
\end{aligned}
\end{equation}
where $P_1$ and $P_2$ are the four-dimensional momenta of $\chi$ and $\chi^\dagger$, respectively; $h^-$ is $\pi^-$ or $K^-$. The hadronic transition matrix element is parameterized the same as that in Eq. (\ref{eq2}) or Eq. (\ref{eq4}), and the form factors are expressed in Eq. (\ref{eq3}) or Eq. (\ref{eq5}).

The transition amplitude receives the contribution from two terms in the effective Lagrangian, and the partial width can be written as
\begin{equation}
\begin{aligned}
\Gamma =\int dPS_3 |g_{p1}\mathcal{T}_1+g_{p3}\mathcal{T}_3|^2=|g_{p1}|^2\widetilde\Gamma_1+|g_{p3}|^2\widetilde\Gamma_3.
\label{eq19}
\end{aligned}
\end{equation}
Here we have defined $\widetilde\Gamma_{1(3)}=\int dPS_3 |\mathcal{T}_{1(3)}|^2$, which are independent of the effective coupling constants. The interference terms are proved to be zero.

\begin{figure}[htb]
	\centering
	\subfigure[$~B\to K$]{\label{}
		\includegraphics[width=0.43\textwidth]{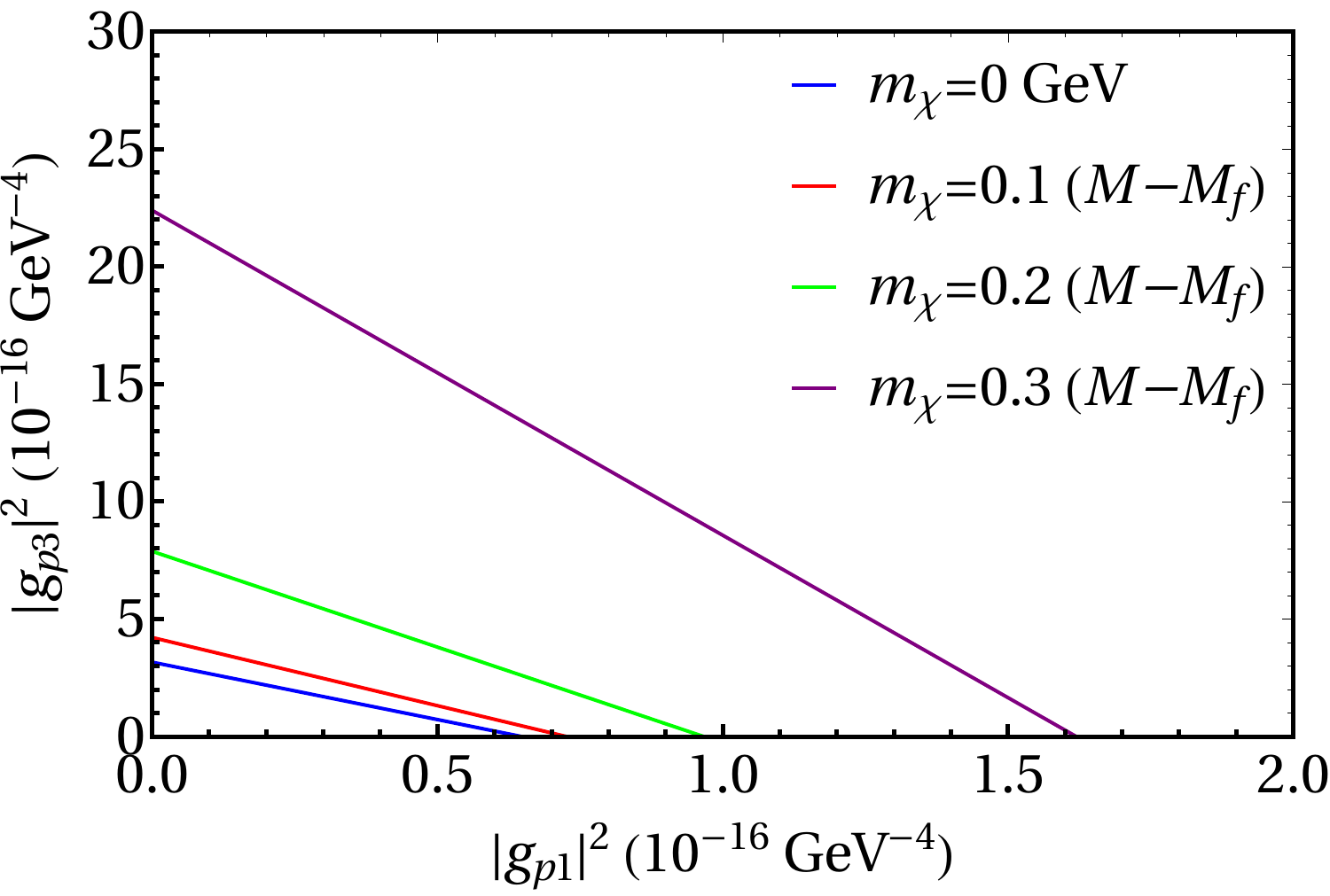}} 
	\hspace{0.4cm}
	\subfigure[$~B\to \pi$]{\label{}
		\includegraphics[width=0.42\textwidth]{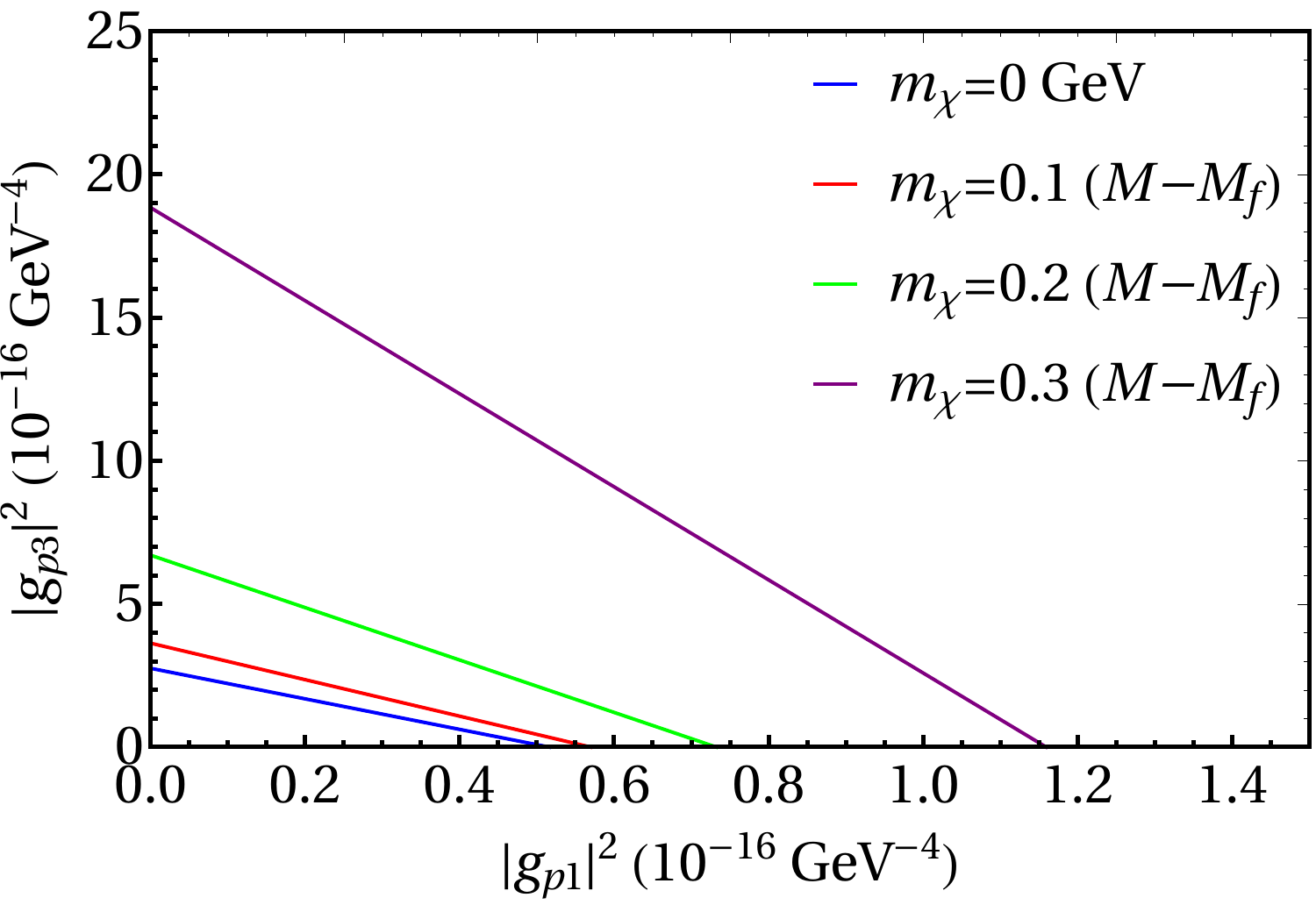}} \\
	\caption{The allowed region of $|g_{p1}|^2$ and $|g_{p3}|^2$ deduced from $B^-\to K^-(\pi^-)\chi\chi^\dagger$.}
	\label{fig5}
\end{figure}

Comparing the theoretical predictions and the experimental upper bound of these channels, we give the possible relations of the modulus square of the effective coupling constants, which are presented in Fig. \ref{fig5}. In this figure, the area below the colored line is allowed experimentally with a specific mass of $\chi$. One can see that as $m_\chi$ increasing, the allowed region gets larger and larger.

If the final meson is a vector, the situation is a little more complicated, because this time the decay processes can be induced by the second, the third, and the fourth operators in the effective Lagrangian. The transition amplitude is 
\begin{equation}
\begin{aligned}
\langle h^{*-}\chi^\dagger\chi|\mathcal L_{2}|B^-\rangle&=\bigg[\frac{g_{p2} m_q}{m_q-m_{q_{_f}}}(P-P_f)_\mu+g_{p4} (P_1-P_2)_\mu\bigg] \langle h^{*-}|(\bar q_{_f} \gamma ^\mu \gamma^5 q) |B^-\rangle \\
&~~~+ g_{p3} (P_1-P_2)_\mu \langle h^{*-}|(\bar q_{_f}\gamma^\mu  q) |B^-\rangle,
\label{eq20}
\end{aligned}
\end{equation}
where $h^{\ast-}$ represents $\rho^-$ or $K^{\ast-}$. Here we need to consider two kinds of hadronic transition matrix elements. $\langle h^{*-}|\bar q_f \gamma^\mu\gamma^5 q |B^-\rangle$ is parameterized the same as Eq. (\ref{eq4}). $\langle h^{*-}|(\bar q_{_f}\gamma^\mu  q) |B^-\rangle $ is expressed as
\begin{equation}
\begin{aligned}
\langle h^{*-}|\bar q_f \gamma^\mu q |B^-\rangle=\frac{2 V(s)}{M+M_f} \varepsilon _{\mu \nu \rho \sigma} \epsilon ^\nu P^\rho P_f^\sigma,
\label{eq21}
\end{aligned}
\end{equation}
where $V(s)$ is expressed by Eq. (\ref{eq5}) and the parameters are given in Table \ref{tab3}.

\begin{figure}[htb]
	\centering
	\subfigure[$~B^-\to K^{*-}\chi\chi^\dagger$]{\label{}
		\includegraphics[width=0.43\textwidth]{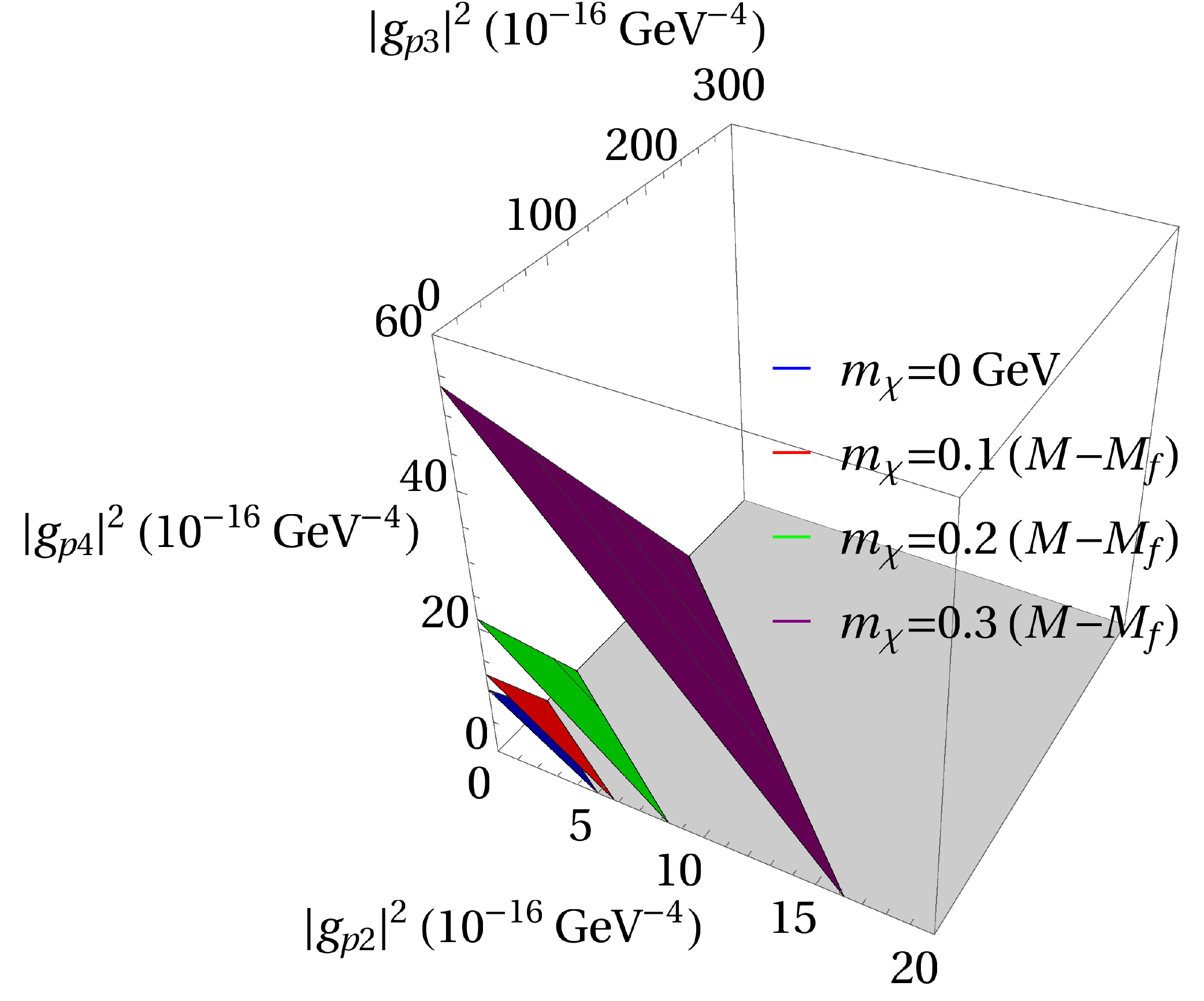}} 
		\hspace{0.3cm}
	\subfigure[$~B^-\to \rho^-\chi\chi^\dagger$]{\label{}
		\includegraphics[width=0.43\textwidth]{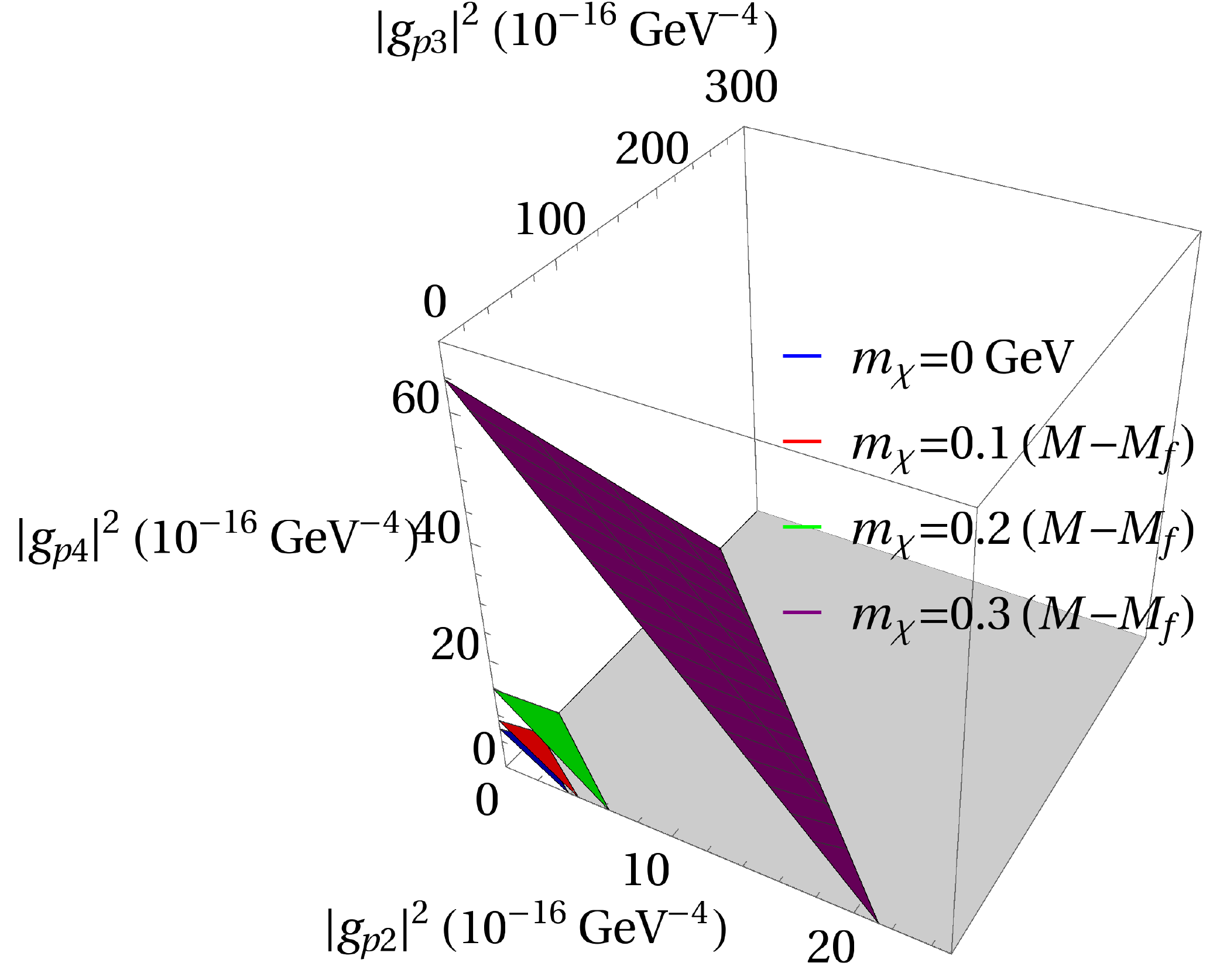}} 
	\caption{The allowed region of $|g_{p2}|^2$, $|g_{p3}|^2$ and $|g_{p4}|^2$ deduced from $B^-\to \rho^-(K^{\ast-})\chi\chi$.}
	\label{fig6}
\end{figure}

The relationship between three effective couplings can be achieved by comparing the theoretical results and the experimental upper limits. Numerical calculation indicates the cross terms can also been neglected. In Fig. \ref{fig6}, we show that the experimentally allowed region is that under the  colored plane which corresponding a specific mass of $\chi$.

\section{The decay modes of the $B_c$ meson}

In the previous section, we have used the results of LCSR to study the FCNC processes of $B$ meson. This method is suitable for the heavy-light state. For the $B_c$ meson, which consists of a heavy quark and a heavy antiquark, we choose the BS method to study its decay processes. The first step is to solve the BS equation which describes the two-body bound state very well. It has the form~\cite{Chang:2014jca}
\begin{equation}
\begin{aligned}
(\slashed p_1-m_1)\chi_P(q)(\slashed p_2+m_2)=\mathrm i\int\frac{\mathrm d^4k}{(2\pi)^4}V(P,k,q)\chi_P(k),
\label{eq22}
\end{aligned}
\end{equation}
where $P$ is the momentum of the meson; $p_1$ and the $p_2$ are the momenta of the quark and antiquark, respectively; $m_1$ and $m_2$ are the masses of the quark and antiquark, respectively; $q$ is the relative momentum between quark and antiquark; $\chi_P(q)$ is the BS wave function; V is the interaction kernel. 

For $B_c$ meson, we can safely make an instantaneous approximation for $V$, that is $V(P,k,q)\approx V(P,k_\perp,q_\perp)$, where $q_\perp= q-\frac{P\cdot q}{\sqrt{P^2}}P$, and the same is for $k_\perp$. By defining the Salpeter wave function $\varphi(q_\perp)={\mathrm i}\int\frac{dq^0}{2\pi}\chi_P(q)$, we reduce Eq. (\ref{eq22}) to the three-dimensional form, which can be solved numerically. $\varphi(q_\perp)$ is constructed from $P$, $q_\perp$, Dirac gamma matrices, and some scalar function of $q_\perp^2$. We take the $0^-$ and $1^-$ states as examples, whose Salpeter wave functions are~\cite{Wang:2009er}
\begin{equation}
\begin{aligned}
&\varphi_{0^-}(q_\perp)=\bigg[ f_1(q_{\perp})+\frac{\slashed{P}}{M}f_2(q_{\perp})+\frac{\slashed{q}_{\perp} }{M}f_3(q_{\perp})+\frac{\slashed{P} \slashed{q}_{\perp}}{M^2}f_4(q_{\perp})\bigg]\gamma_{5},\\
&\varphi _{1^-}(q_\perp)
= (q_{\perp} \cdot \epsilon) \bigg[g_1(q_{\perp})+\frac{\slashed{P}}{M}g_2(q_{\perp})+\frac{q_{\perp} }{M}g_3(q_{\perp})+\frac{\slashed{P} \slashed{q}_{\perp} }{M}g_4(q_{\perp})\bigg] \\
&~~~~~~~~~~~~~~+ M\bigg[g_5(q_{\perp})+\frac{\slashed{P}}{M}g_6(q_{\perp})+\frac{q_{\perp} }{M_{f}}g_7(q_{\perp})+\frac{\slashed{P}\slashed{q}_{\perp} }{M^2}g_8(q_{\perp})\bigg]\slashed{\epsilon}.
\label{eq23}
\end{aligned}
\end{equation}

In Mandelstam formalism, the hadronic transition matrix element can be expressed as the overlap integral of the BS wave functions of the initial and final mesons. With the instantaneous approximation, it can be reduced to the overlap integral of Salpeter wave functions. To make the calculation simple, we just keep the positive energy parts of the wave functions which give the main contribution. The transition amplitude is~\cite{Mandelstam:1955sd}
\begin{equation}
\begin{aligned}
\langle h^-|\bar q_1\Gamma^\xi b|B_c^-\rangle 
&= \int\frac{d^3 q}{(2\pi)^3} {\rm Tr}\left[\frac{\slashed P}{M}\overline\varphi_{P_f}^{++}(q_{f\perp})\Gamma^\xi\varphi_P^{++}(q_\perp)\right],
\label{eq24}
\end{aligned}
\end{equation}
where $\varphi^{++}(q_\perp)=\Lambda^{+}_1\frac{\slashed P}{M}\varphi(q_\perp)\frac{\slashed P}{M}\Lambda^{+}_2$. Here we have used the definition of the positive energy projector operator $\Lambda_i^+=\frac{1}{2\omega_i}\bigg[\frac{\slashed P}{M}\omega_i- (-1)^{i}(\slashed q_\perp + m_i)\bigg]$ with $i=1,2$.

\subsection{The SM backgroud}

\begin{figure}[htb]
	\centering
	\subfigure[]{\label{}
		\includegraphics[width=0.355\textwidth]{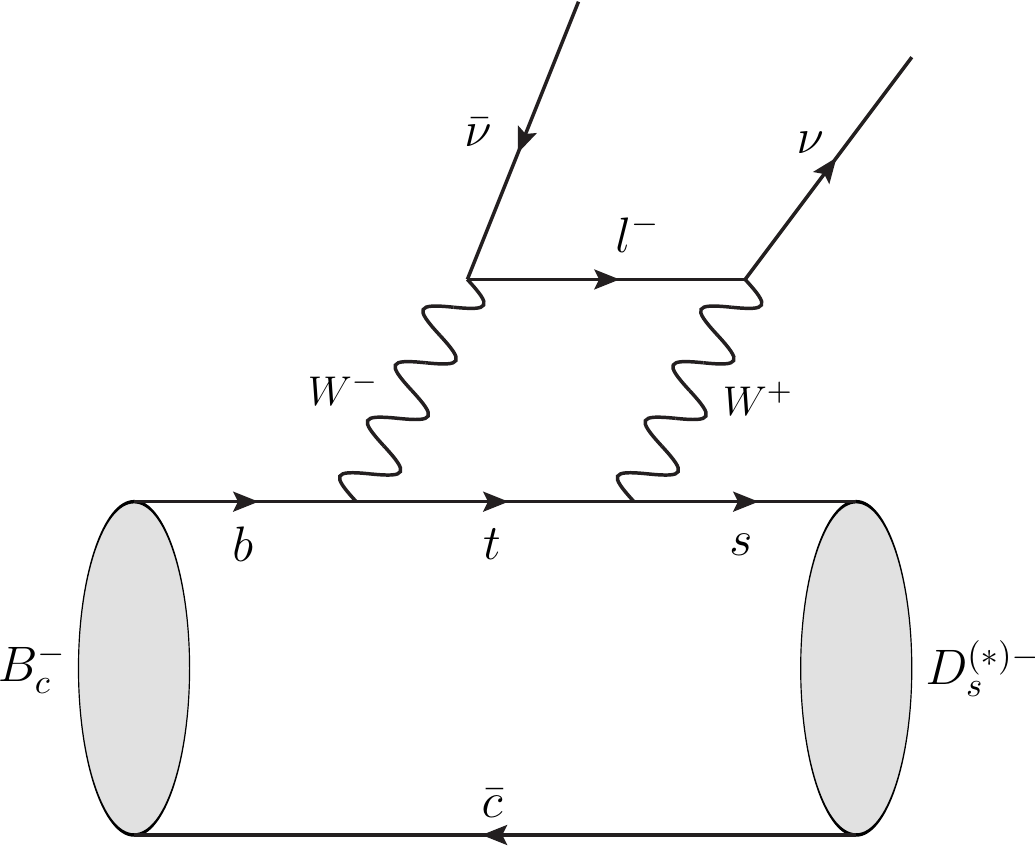}}
	\hspace{0.5cm}
	\subfigure[]{\label{}
		\includegraphics[width=0.355\textwidth]{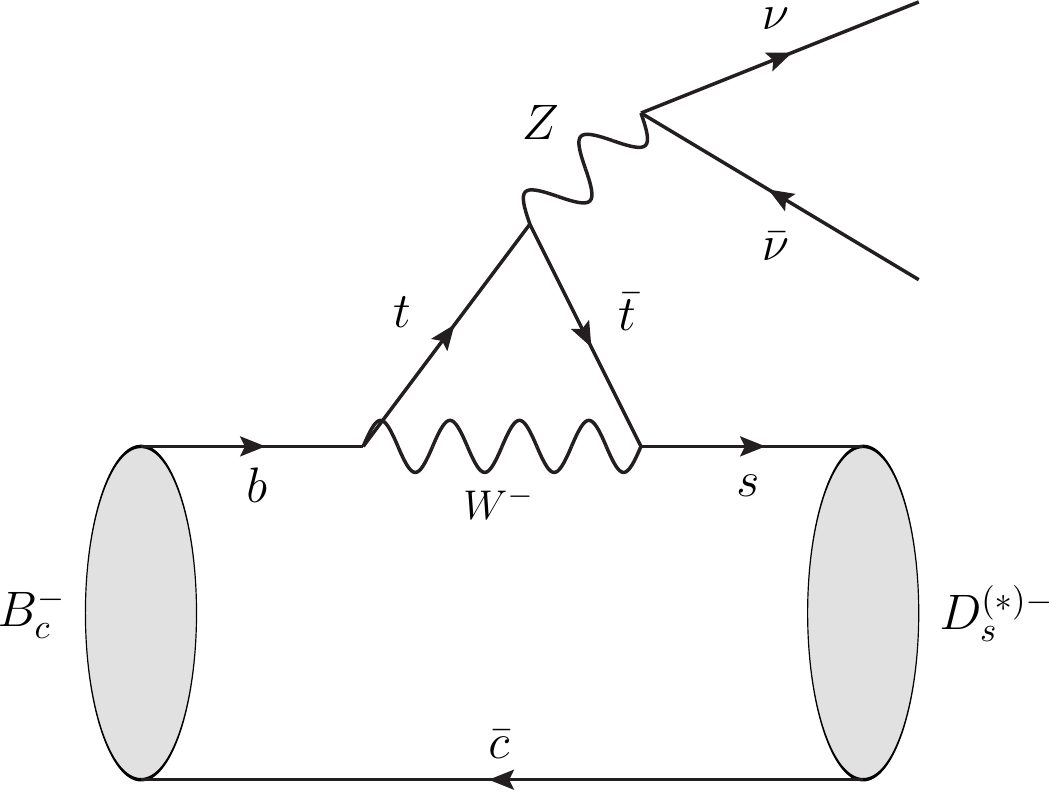}}
	\caption{Feynman diagrams for the process $B_c^-\rightarrow D_s^{(\ast)-}\nu\bar \nu$.}
	\label{fig7}
\end{figure}

In the Standard Model, the missing energy in the decay processes $B_c^-\rightarrow D_s^{(\ast)-}+\slashed E$ is carried by the (anti)neutrino. The corresponding Feynman diagrams are presented in Fig. \ref{fig7}. 
It can be described by an effective Lagrangian
\begin{equation}
\begin{aligned}
\mathcal L_{3} = \frac{4G_F}{\sqrt{2}}\frac{\alpha}{2\pi\sin^2\theta_W}\sum_{l=e, \mu, \tau}\sum_{q=u, c, t}V_{bq}V_{sq}X^l(x_q)(\bar s_{_L}\gamma^\mu b_{_{L}})(\bar\nu_{_{lL}}\gamma_\mu\nu_{_{lL}}),
\label{eq25}
\end{aligned}
\end{equation}
where $G_F$ is the Fermi coupling constant; $\alpha$ is the fine structure constant; $\theta_W$ is the Weinberg angle; $V_{q_1q_2}$ is the CKM matrix element; $X^l(x_t)$ is the Inami-Lim function~\cite{Inami:1980fz}, which has the form
\begin{equation}
\begin{aligned}
X^l(x_t) = \frac{x_t}{8}\left[\frac{x_t+2}{x_t-1}+\frac{3(x_t-2)}{(x_t-1)^2}\ln x_t\right],
\label{eq26}
\end{aligned}
\end{equation}
with $x_t=m_t^2/M_W^2$.

The transition amplitude is
\begin{equation}
\begin{aligned}
\langle D_s^{(\ast)-}\nu_{_l}\bar\nu_{_l}|\mathcal L_{3}|B_c^-\rangle &=\frac{\sqrt{2}G_F\alpha}{4\pi\sin^2\theta_W}V_{bt}V_{st}X^l(x_t)\langle D_s^{(\ast)-}|\bar s\gamma^\mu(1-\gamma^5)b|B_c^-\rangle\\
&~~~\times \bar u_{\nu_l}\gamma_\mu(1-\gamma^5)v_{\nu_l}.
\label{eq27}
\end{aligned}
\end{equation}
The hadronic transition matrix element is calculated by Eq. (\ref{eq24}). The branching fraction is achieved by finishing the three-body phase space integral, which is presented in Table \ref{tab4} to compare with the results of other models. The errors come from varying the parameters in our model by $\pm5\%$.

\begin{figure}[htb]
	\centering
	\subfigure[]{\label{}
		\includegraphics[width=0.355\textwidth]{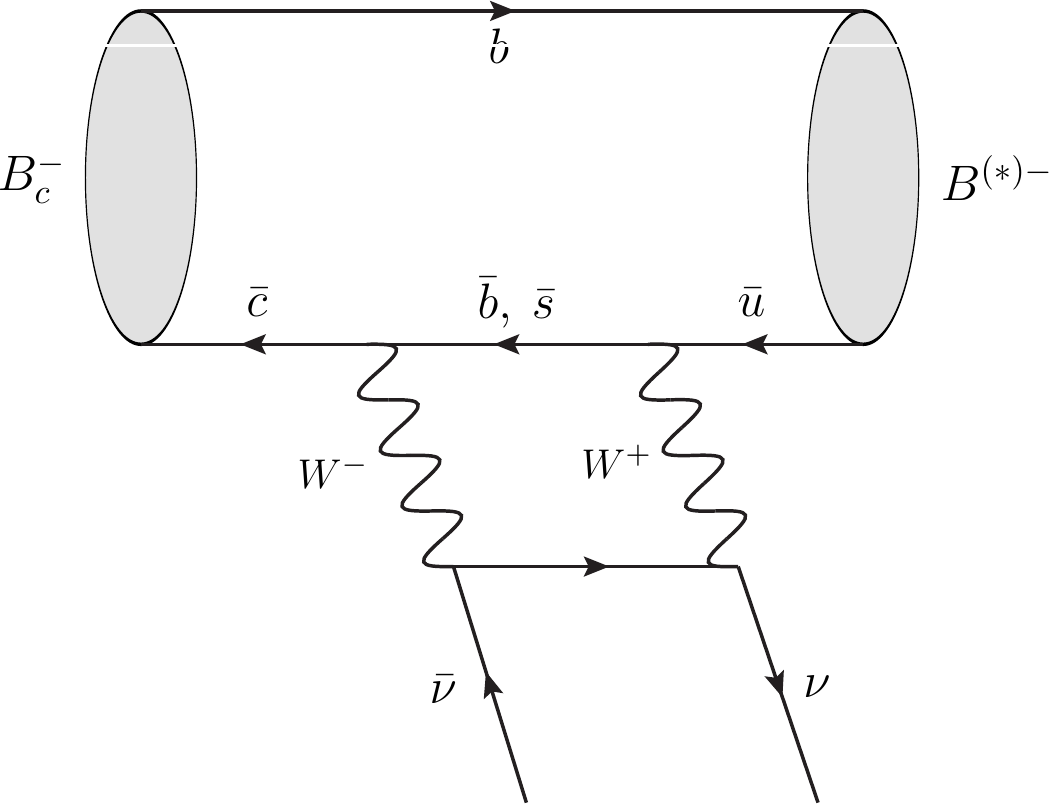}}
	\hspace{0.5cm}
	\subfigure[]{\label{}
		\includegraphics[width=0.355\textwidth]{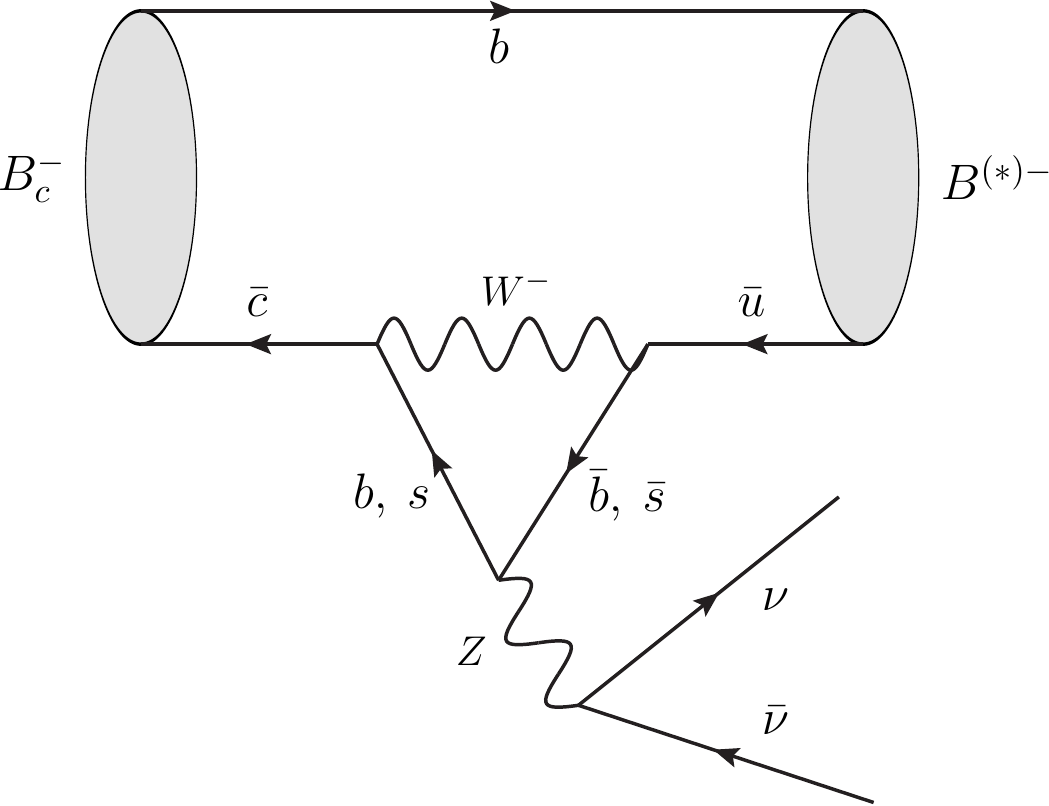}}
	\caption{Feynman diagrams for the process $B_c^-\rightarrow B^{(\ast)-}\nu\bar \nu$.}
	\label{fig8}
\end{figure}

There are also the $B_c\to B_u\nu\bar\nu$ processes, which is induced by $c\rightarrow u$ at the quark level. The Feynman diagrams for such channels are given in Fig. \ref{fig8}. 
The corresponding effective Lagrangian is 
\begin{equation}
\begin{aligned}
\mathcal L_{4} = \frac{4G_F}{\sqrt{2}}\frac{\alpha}{2\pi\sin^2\theta_W}\sum_{l=e, \mu, \tau}\sum_{q=s,b}V_{cq}V_{uq}X^l(x_q)(\bar u_{_L}\gamma^\mu c_{_{L}})(\bar\nu_{_{lL}}\gamma_\mu\nu_{_{lL}}),
\label{eq28}
\end{aligned}
\end{equation}
where
\begin{equation}
\sum_qV^\ast_{cq}V_{uq}X^l(x_q)=V^\ast_{cs}V_{us}X^l(x_s)+V^\ast_{cb}V_{ub}X^l(x_b).
\label{eq29}
\end{equation}
In the above equation, we have defined $X^l(x_q)=\bar D(x_q, y_l)/2$, and the Inami-Lim function $\bar D(x_q, y_l)$ is expressed as~\cite{Inami:1980fz}
\begin{equation}
\begin{aligned}
\bar D(x_q, y_l) &= \frac{1}{8}\frac{x_qy_l}{x_q-y_l}\left(\frac{y_l-4}{y_l-1}\right)^2\ln y_l + \frac{1}{8}\left[\frac{x_q}{y_l-x_q}\left(\frac{x_q-4}{x_q-1}\right)^2+1+\frac{3}{(x_q-1)^2}\right]\\
&~~~\times x_q\ln x_q +\frac{x_q}{4}-\frac{3}{8}\left(1+\frac{3}{y_l-1}\right)\frac{x_q}{x_q-1},
\label{eq30}
\end{aligned}
\end{equation}
where we have used $x_q=m_q^2/M_W^2$ and $y_l=m_l^2/M_W^2$.
The transition amplitude has the form
\begin{equation}
\begin{aligned}
\langle B^{(\ast)-}\nu_{_l}\bar\nu_{_l}|\mathcal L_{4}|B_c^-\rangle &=\frac{\sqrt{2}G_F\alpha}{4\pi\sin^2\theta_W}\sum_qV^\ast_{cq}V_{uq}X^l(x_q)\langle B^{(\ast)-}|\bar c\gamma^\mu(1-\gamma^5)u|B_c^-\rangle\\
&~~~\times \bar u_{\nu_l}\gamma_\mu(1-\gamma^5)v_{\nu_l}.
\label{eq31}
\end{aligned}
\end{equation}
\begin{table}[htb]
	\caption{ The branching fractions of the rare semileptonic $B_c$ decays (in units of $10^{-8}$).}
	\vspace{0.2cm}
	\label{}
	\setlength{\tabcolsep}{0.2cm}
	\centering
	\begin{tabular*}{\textwidth}{@{}@{\extracolsep{\fill}}cccccc}
		\hline\hline
		Mode&Ours&Ebert~\cite{Ebert:2010dv}&Choi~\cite{Choi:2010ha}&Geng~\cite{Geng:2001vy}&pQCD~\cite{Wang:2014yia}\\
		\hline
		{\phantom{\Large{l}}}\raisebox{+.2cm}{\phantom{\Large{j}}}
		$B_c\rightarrow D_s\bar\nu\nu$&$43.1^{+6.7}_{-6.0}$&$65$&$39$&$92$& $129$\\ 
		{\phantom{\Large{l}}}\raisebox{+.2cm}{\phantom{\Large{j}}}
		$B_c\rightarrow D_s^\ast\bar\nu\nu$&$250^{+16}_{-15}$ &$135$&&$312$&$404$\\
		{\phantom{\Large{l}}}\raisebox{+.2cm}{\phantom{\Large{j}}}
		$B_c\rightarrow D_d\bar\nu\nu$&$1.07^{+0.18}_{-0.19}$ &$2.16$&$1.31$&$2.77$&$3.13$\\
		{\phantom{\Large{l}}}\raisebox{+.2cm}{\phantom{\Large{j}}}
		$B_c\rightarrow D_d^\ast\bar\nu\nu$&$7.32^{+0.96}_{-0.91}$ &$5.12$&&$7.64$&$11$\\
		{\phantom{\Large{l}}}\raisebox{+.2cm}{\phantom{\Large{j}}}
		$B_c\rightarrow B_u\bar\nu\nu$& $5.15^{+1.21}_{-1.19}\times10^{-7}$&&&&\\
		{\phantom{\Large{l}}}\raisebox{+.2cm}{\phantom{\Large{j}}}
		$B_c\rightarrow B_u^\ast\bar\nu\nu$& $1.11^{+0.14}_{-0.19}\times10^{-6}$&&&&\\
		\hline\hline
	\label{tab4}
	\end{tabular*}
\end{table}

\subsection{The $B_c\to P(V)\chi\chi$ process}
These processes are induced by the same Lagrangian in Eq. (\ref{eq1}). For the final meson being a pseudoscalar ($P$) or a vector ($V$), the decay amplitudes are
\begin{equation}
\begin{aligned}
\langle h^-\chi\chi|\mathcal L_{1}|B_c^-\rangle&=2 g_{s1} m_q \langle h^-|(q_{_f}q) |B_c^-\rangle
\label{eq32}
\end{aligned}
\end{equation}
and 
\begin{equation}
\begin{aligned}
\langle h^{\ast-}\chi\chi|\mathcal L_{1}|B_c^-\rangle&=2 g_{s2} m_q \langle h^{\ast-}|(q_{_f}\gamma^5 q) |B_c^-\rangle,
\label{eq33}
\end{aligned}
\end{equation}
respectively, where $h^{(\ast)}$ can be $D^{(\ast)}$, $D_s^{(\ast)}$, or $B^{(\ast)}$, and the hadronic transition matrix elements are calculated with Eq. (\ref{eq24}). By finishing the three-body phase space integral, we get the decay widths expressed as the product of the squared effective coupling constant $|g_{si}|^2$ and the quantity $\widetilde\Gamma_i$.

\begin{figure}[htb]
	\centering
	\subfigure[~$b\to s(d)$]{\label{fig9a}
		\includegraphics[width=0.43\textwidth]{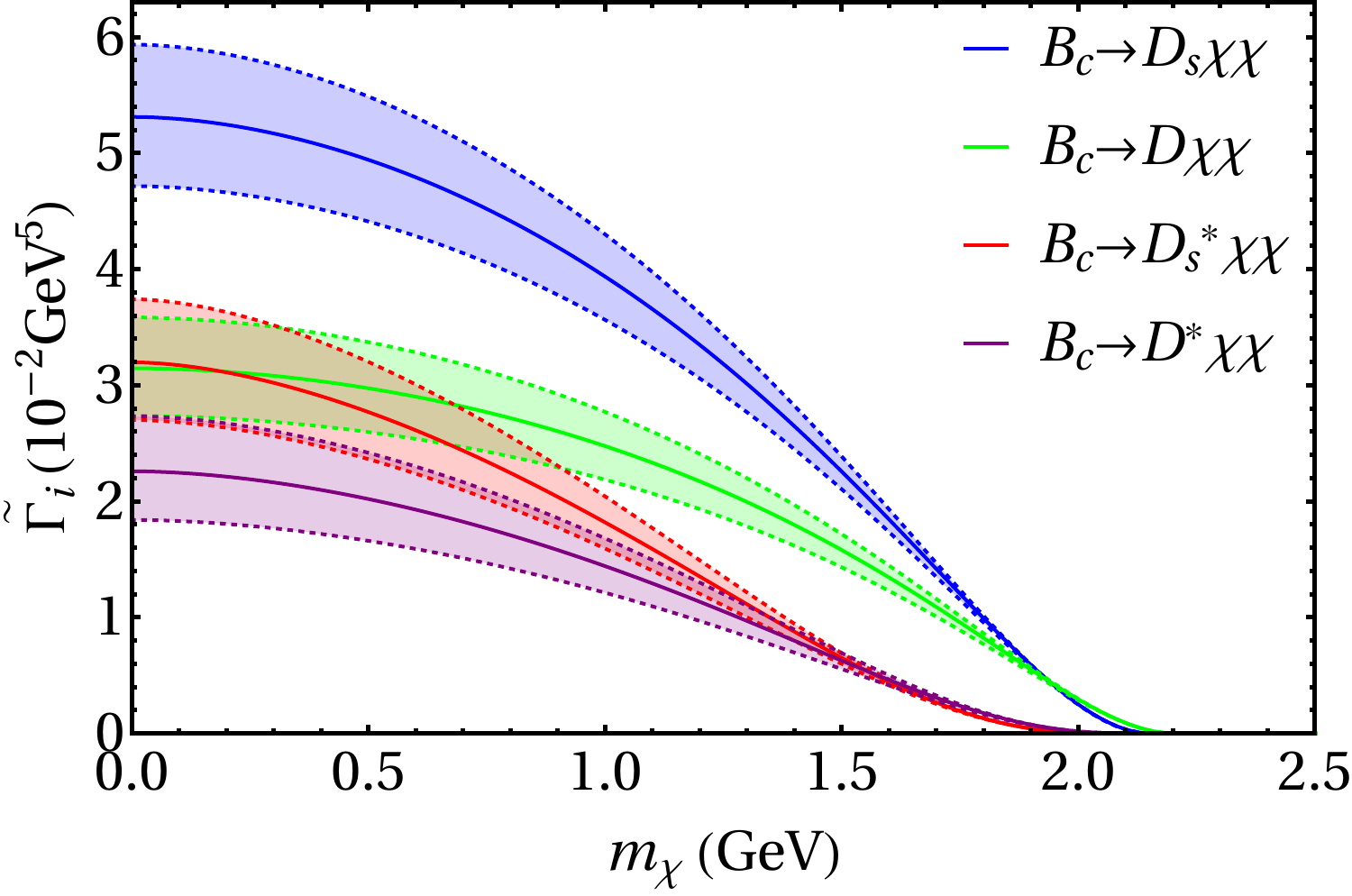}} 
		\hspace{0.5cm}
	\subfigure[~$c\to u$]{\label{fig9b}
		\includegraphics[width=0.43\textwidth]{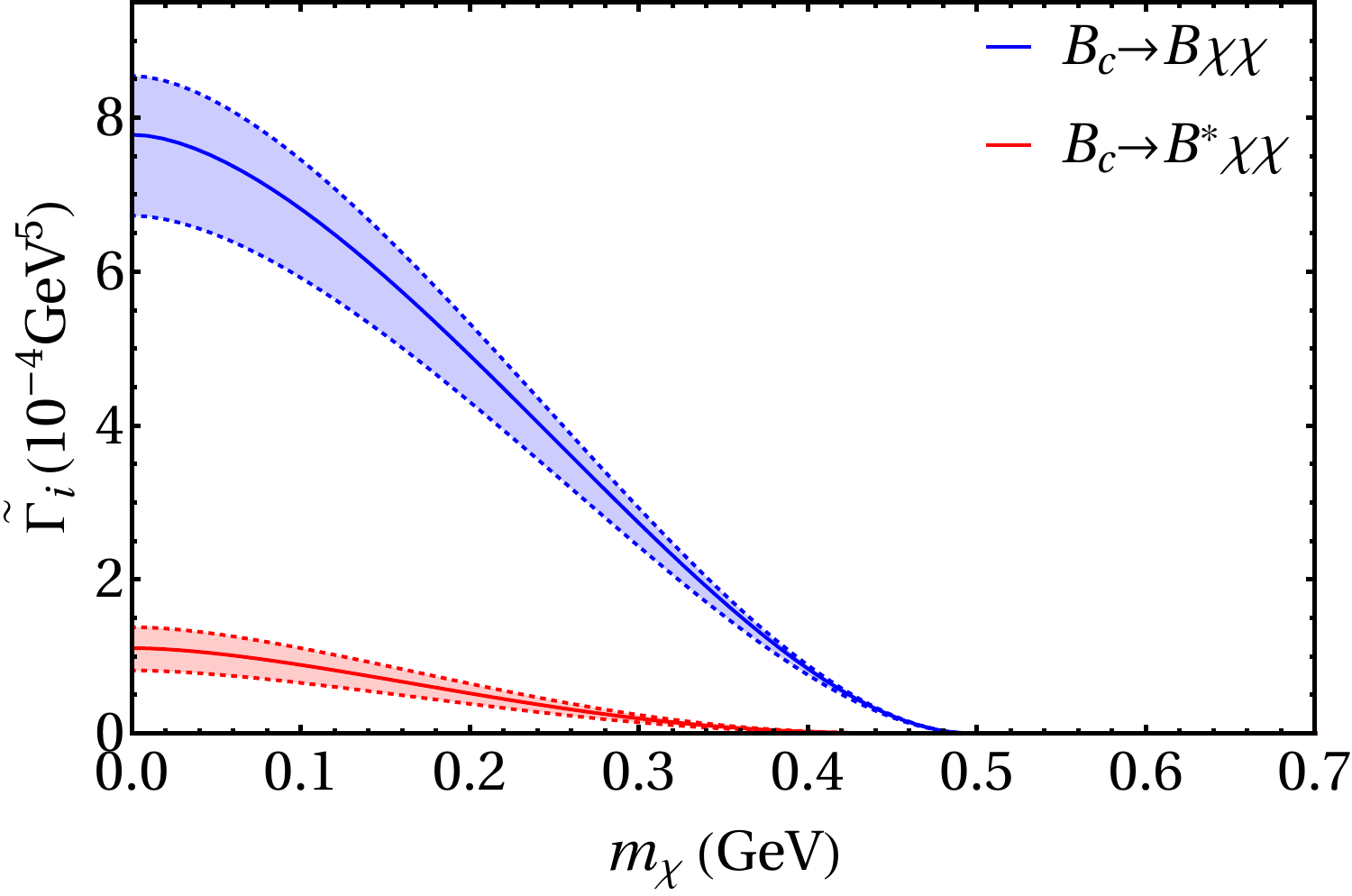}} 
	\caption{The quantity $\widetilde\Gamma_i$ changes with $m_\chi$ in the $B_c^-\to P(V)\chi\chi$ process. The shadows represent the errors estimated by varying the parameters in our model by $\pm5\%$.}
	\label{fig9}
\end{figure}

$\widetilde\Gamma_i$ is independent of the coupling constants, and can be calculated by taking a specific value of $m_\chi$. In Fig. \ref{fig9}, we plot them as functions of $m_\chi$.   One can see that they all decrease when $m_\chi$ gets larger, because the phase space gets smaller. With the same value of $m_\chi$, $\widetilde\Gamma_1$ from the $B_c\to P\chi\chi$ channels is larger than $\widetilde\Gamma_2$ from the $B_c\to V\chi\chi$ channels due to the different effective vertex in the amplitude. For $\widetilde\Gamma_{1}$, the $B_c\to D_s$ channel gives a larger result than that of the $B_c\to D$ channel when $m_\chi<1.9$ GeV. When $m_\chi$ gets even larger, the phase space suppression will be important. For $\widetilde\Gamma_2$, this turning point is about 1.53 GeV. We also notice that $\widetilde\Gamma_i$ of the $c\to u$ processes is two orders of magnitude less than that of the $b\to d(s)$ processes. This comes from both the smaller phase space and smaller $m_q$ for the former case.

\begin{figure}[htb]
	\centering
	\subfigure[$~B_c\to P$]{
		\label{fig10a}
		\includegraphics[width=0.43\textwidth]{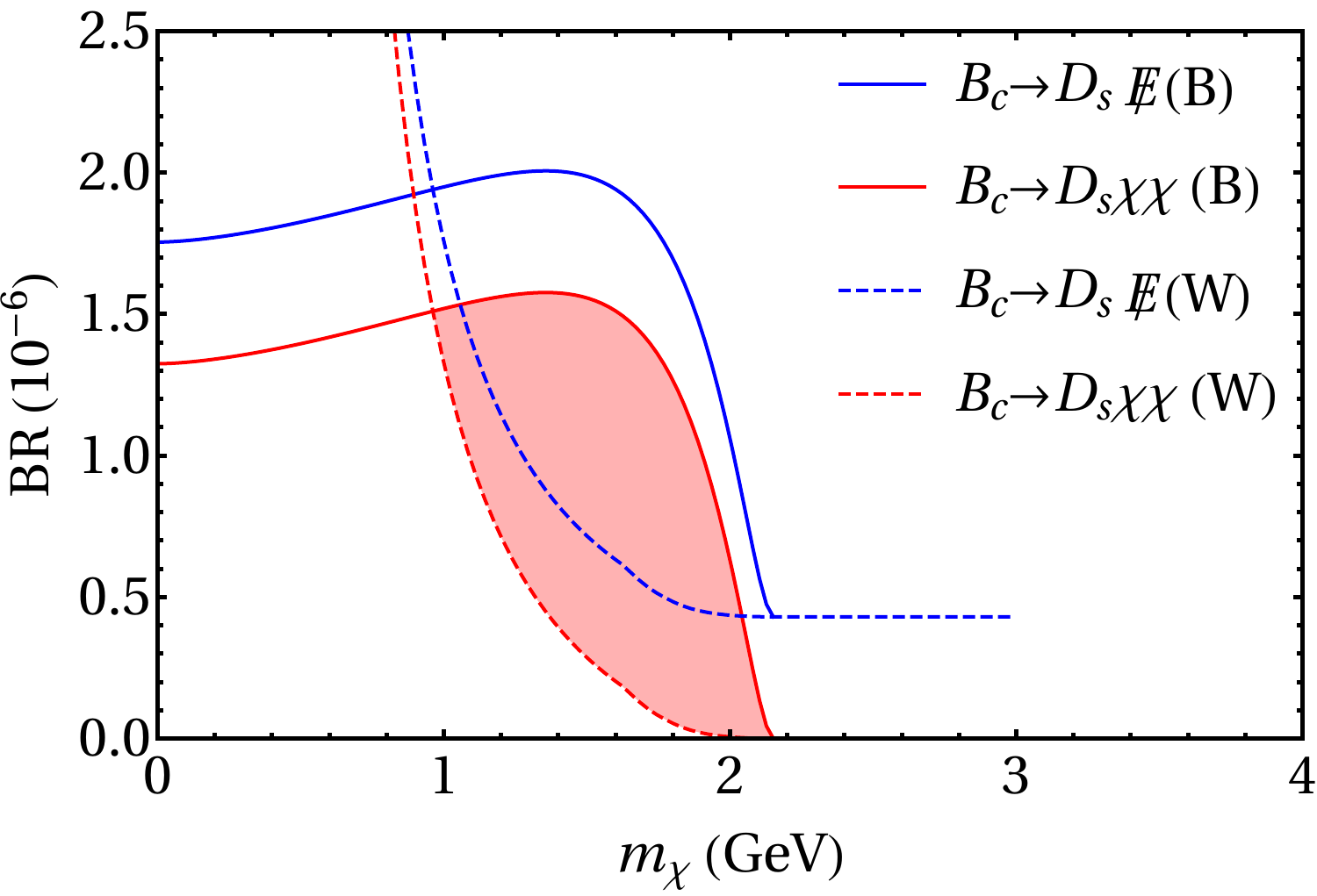}}
		\hspace{0.5cm}
	\subfigure[$~B_c\to P$]{
		\label{fig10b}
		\includegraphics[width=0.43\textwidth]{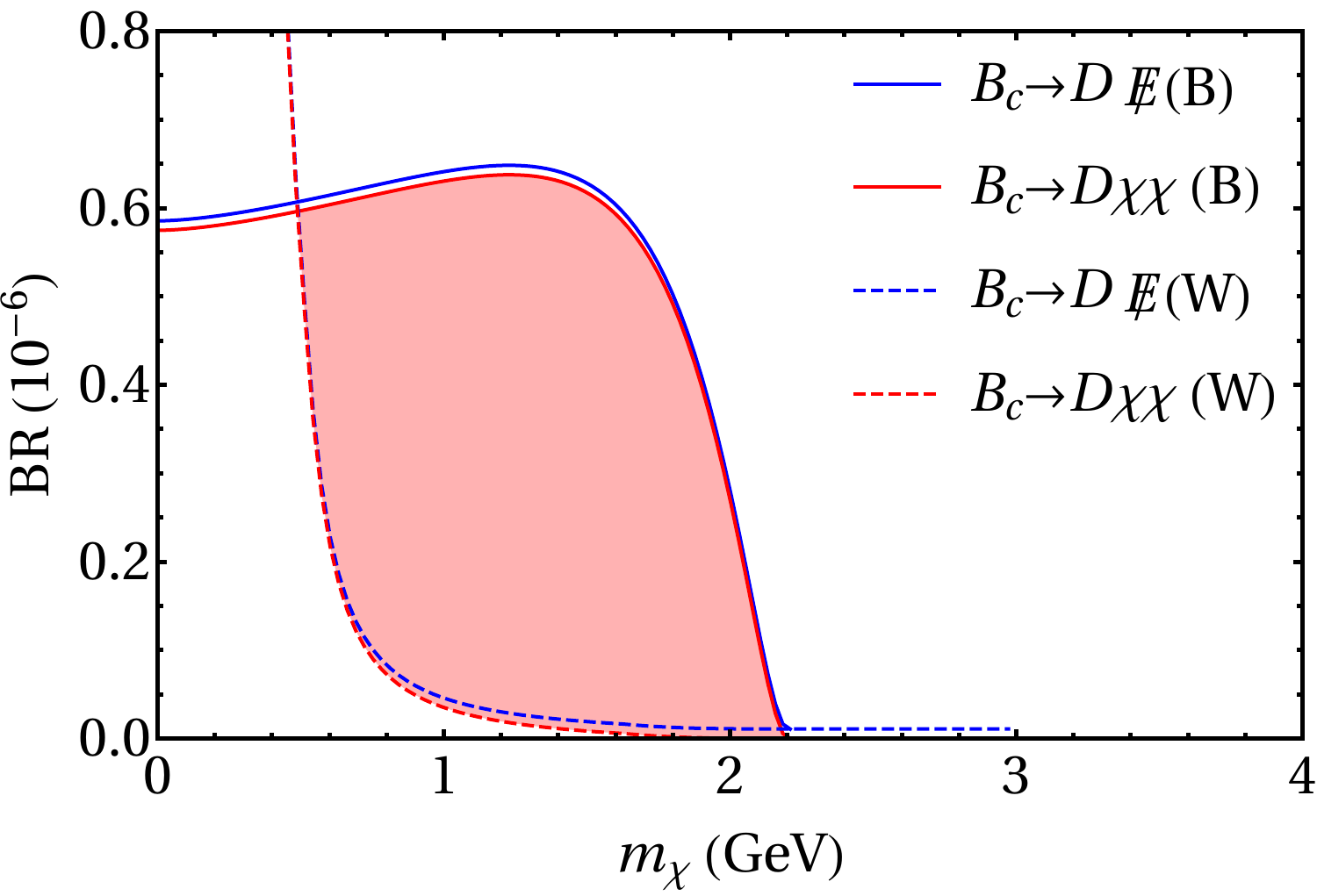}}\\
	\subfigure[$~B_c\to V$]{
		\label{fig10c}
		\includegraphics[width=0.43\textwidth]{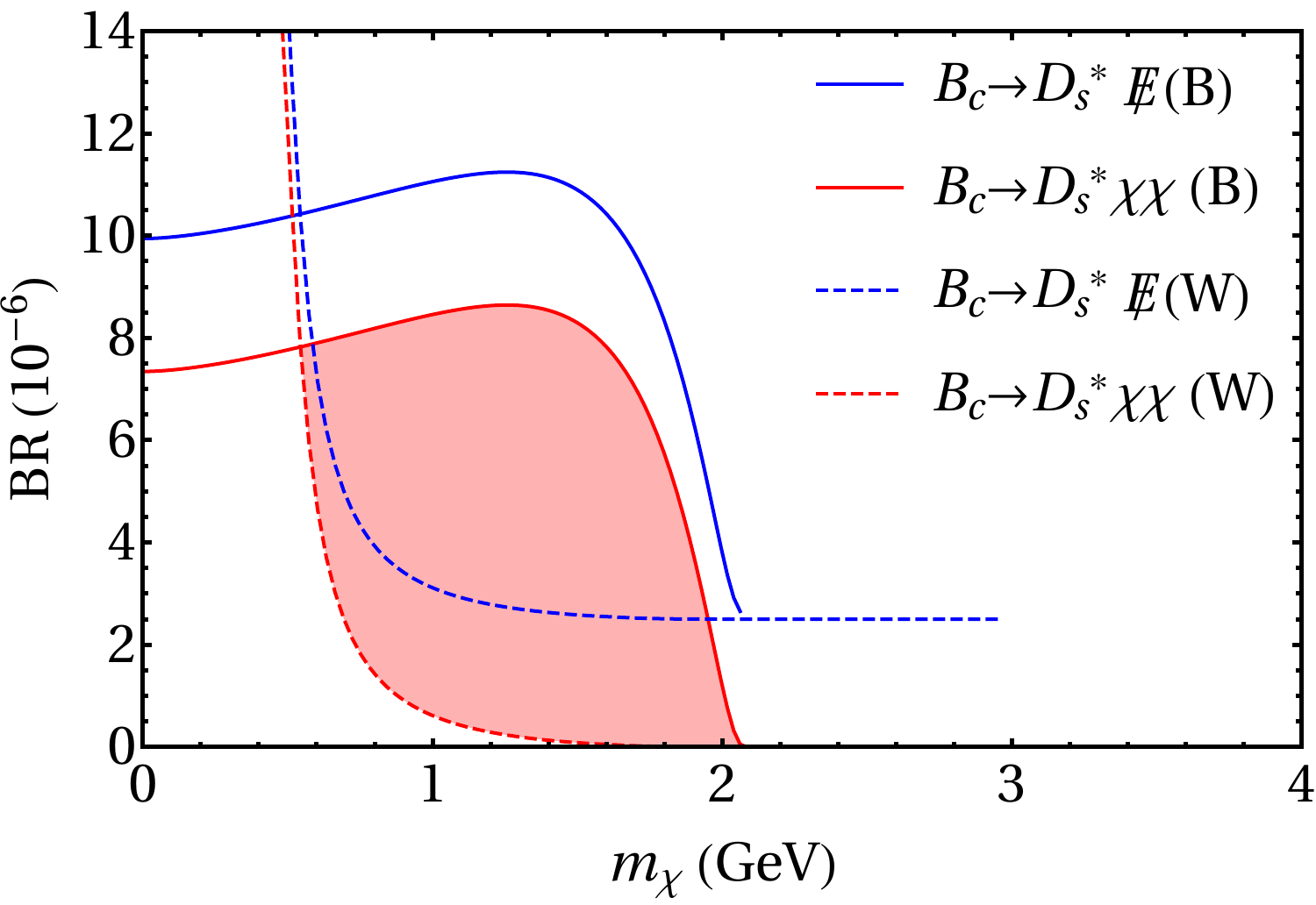}}
			\hspace{0.5cm}
	\subfigure[$~B_c\to V$]{
		\label{fig10d}
		\includegraphics[width=0.43\textwidth]{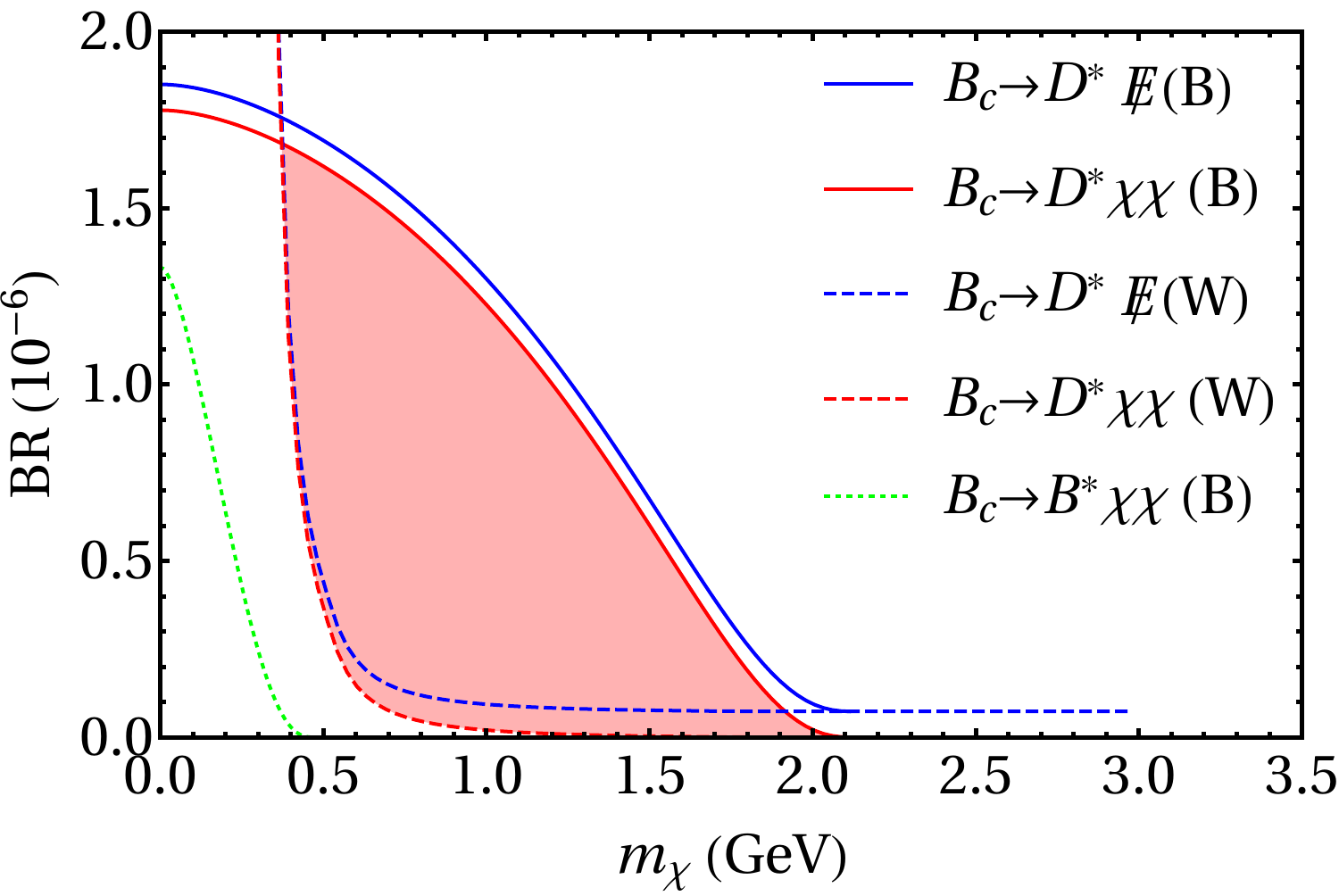}}\\
	\caption{Branching ratios of $B_c\to P(V)\chi\chi$.}
	\label{fig10}
\end{figure}

The upper limits of the squared coupling constants $|g_{s_i}|^2$ with different values of $m_\chi$ have been given in Fig. \ref{fig3}. Combining the results in Fig. \ref{fig3} and Fig. \ref{fig9}, we can make predictions of the upper limits of the branching ratios of the $B_c\to P(V)\chi\chi$ channels. In Fig. \ref{fig10} we present the results which are represented by the red solid lines. For the $B_c\to D\chi\chi$ and $B_c\to D_s\chi\chi$ processes, we use the upper limits extracted from the $B\to\pi\chi\chi$ and $B\to K\chi\chi$ channels, respectively (see Fig. \ref{fig3a}). For the $B_c\to D_s^\ast\chi\chi$ channel, the result of $B\to K^\ast\chi\chi$ is used. For the $B_c\to D^\ast\chi\chi$ case, there are two processes available to set the upper limits, namely $B\to\rho\chi\chi$ and $B\to\chi\chi$. The later one gives more stringent constraint, which is applied here. As there is no experimental result now for the the $D\to\pi+\slashed E$, we cannot give the constraints for the $B_c\to B\chi\chi$ channel. The $|g_{s2}|^2$ extracted from the $D^0\to\chi\chi$ is used to set the upper limit for the branching ratio of the $B_c\to B^\ast\chi\chi$ channel. 

The upper limits of the branching ratios of $B_c\to P(V)\chi\chi$ channels are of the order of $10^{-6}$ when $m_\chi$ is not close its maximum value.  For $B_c\to D\chi\chi$ and $B_c\to D_s^{(\ast)}\chi\chi$, a specific feature appears. When $m_\chi$ is less than about 1.5 GeV, the branching ratios increases slowly with $m_\chi$; after that, the branching ratios decreases rapidly to zero. It is the result of a combination of the increasing $|g_{si}|^2$ and decreasing $\widetilde\Gamma_i$. One notices that in Fig. \ref{fig9a} the $\widetilde\Gamma$ of the $D_{(s)}^\ast$ case is smaller than that of the $D_{(s)}$ case, however, the branching ratios of the former are several times larger than that of the later, because the experimental upper bound of $B$ and $D$ mesons in Table \ref{tab1} are different. We also predict the upper limits of the branching ratios of $B_c\to P(V)\slashed E$ by assuming that it equals to the sum of the branching ratios of $B_c\to P(V)\chi\chi$ and $B_c\to P(V)\nu\bar\nu$. The results are presented in Fig. \ref{fig10} by the blue solid lines. For the $B_c\to D_s^{(\ast)}\slashed E$ modes, the upper limits of the branching ratios deviate obviously from that of $B_c\to D_s^{(\ast)}\chi\chi$, because the later has the same order of magnitude as that of the SM backgroud when $m_\chi$ is not quite large. For the  $B_c\to D^{(\ast)}\slashed E$ and $B_c\to B^\ast\slashed E$ modes, the upper limits of their branching ratios are very close to that of the corresponding $\chi\chi$ channel. This is because the SM background is small, for example, the branching ratios of $B_c\to B^{(*)}\nu\bar\nu$ is of the order of $10^{-14}$. This provides a way to test our results. If the future experiments find a quit large branching ratio of such channels compared with the SM prediction, it definitely indicates the existence of some new physics. We also present the lower bounds of the branching ratio which come from the constraints of the relic density if $\chi$ is a dark matter. They are represented by the dashed lines in Fig.~\ref{fig10}. The shadow areas are the allowed regions of the branching ratios. It should be pointed that these lower bounds are model-dependent.

\subsection{The $B_c\to P(V)\chi\chi^\dagger$ process}

If $\chi$ is a pseudoscalar, the Lagrangian $\mathcal L_2$ is applied. When the final meson is a pseudoscalar, the transition amplitude has the form
\begin{equation}
\begin{aligned}
\langle h^-\chi\chi^\dagger|\mathcal L_{2}|B^-\rangle&= g_{p1} m_q \langle h^-|(q_{_f}q) |B^-\rangle+g_{p3}(P_1-P_2)_\mu \langle h^-|(q_{_f}\gamma^\mu q) |B^-\rangle,
\label{eq34}
\end{aligned}
\end{equation}
and for the vector meson case, the transition amplitude can be written as
\begin{equation}
\begin{aligned}
\langle h^{\ast-}\chi\chi^\dagger|\mathcal L_{2}|B^-\rangle&= g_{p2} m_q \langle h^{\ast-}|(q_{_f}\gamma^5q) |B^-\rangle+g_{p3}(P_1-P_2)_\mu \langle h^{\ast-}|(q_{_f}\gamma^\mu q) |B^-\rangle\\
&~~~+g_{p4}(P_1-P_2)_\mu \langle h^{\ast-}|(q_{_f}\gamma^\mu\gamma^5 q) |B^-\rangle.
\label{eq35}
\end{aligned}
\end{equation}
The hadronic transition matrix elements are also calculated with Eq. (\ref{eq24}). But this situation is more complicated, because there are two or three operators contribute to the decay. Similar to the subsection II.B, we will neglect the cross terms, and keep the ones proportional to $|g_{pi}|^2$. These terms are named as $\widetilde\Gamma_i$, which are independent of the effective coupling constants but depend on the mass of the light dark matter.

To compare the contribution of different terms, we calculate them by giving a specific value of $m_\chi$. The results are presented in Fig. \ref{fig12}. We can see all the $\widetilde\Gamma_is$ are decreasing with $m_\chi$, which is due to the suppression of phase space. When the final meson is $D$ or $D_s$, $\widetilde\Gamma_1$ is large than $\widetilde\Gamma_3$, while for $B$, the situation is very different.  When the final meson is a vector, $\widetilde\Gamma_2$ and $\widetilde\Gamma_4$ are close to each other, but both larger than $\widetilde\Gamma_4$. The approach we get the upper limits of the decay width is as follows. When $m_\chi$ is given, there is an experimental allowed region for the effective coupling constants which are presented in Fig. \ref{fig5} and Fig. \ref{fig6}. So we scan the parameter space to get the maximum value of the partial decay width. The results are given in Fig. \ref{fig12}. For the $B_c\to B^{(\ast)}\chi\chi^\dagger$ channels, as there are no constraints for the effective coupling constants available now, so the upper limits of their branching ratios cannot be calculated.

\begin{figure}[htb]
	\centering
	\subfigure[$~B_c^-\to D_s^-$]{
		\includegraphics[width=0.43\textwidth]{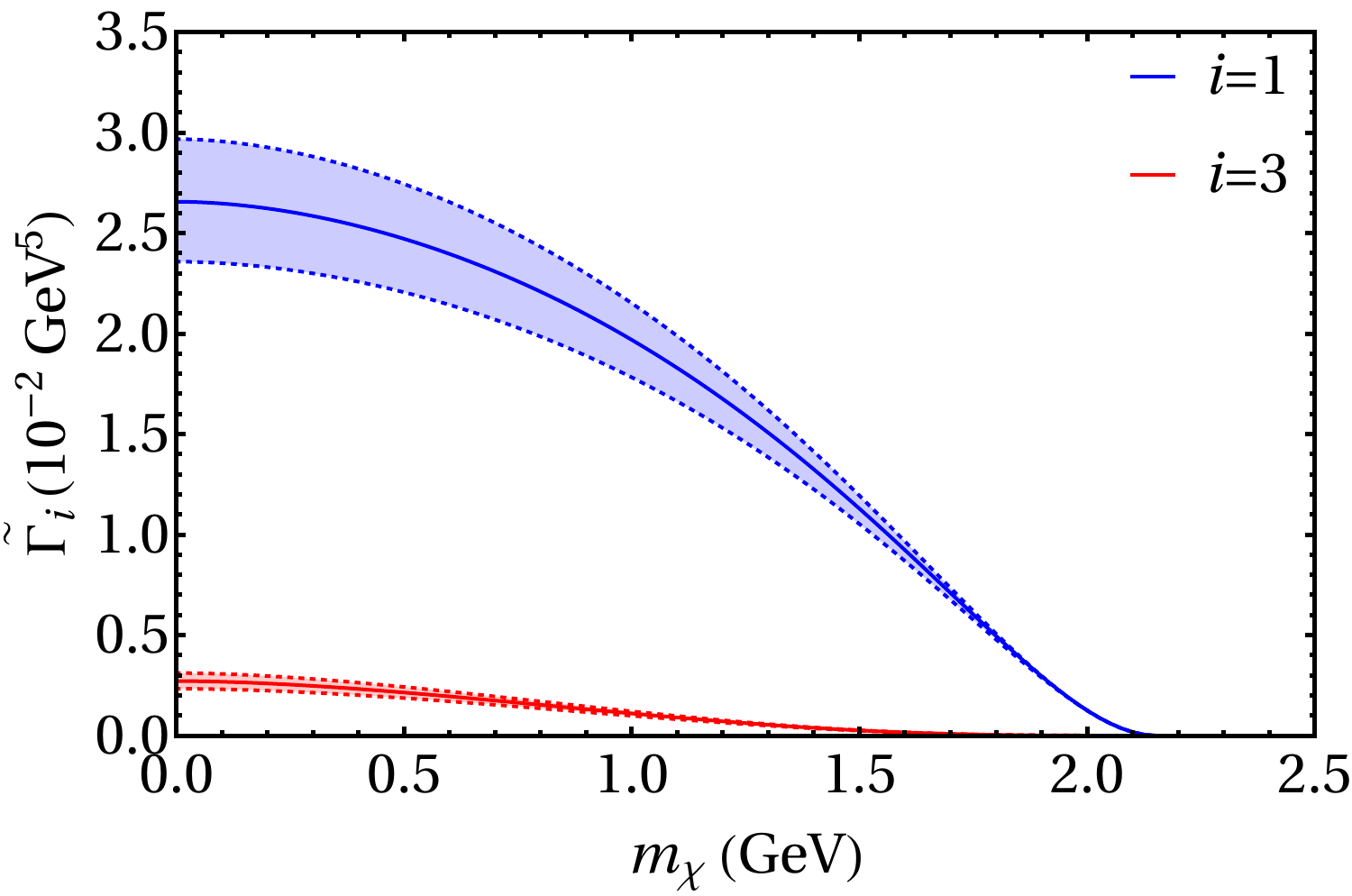}}
	\hspace{0.4cm}
	\subfigure[$~B_c^-\to D_s^{*-}$]{
		\includegraphics[width=0.43\textwidth]{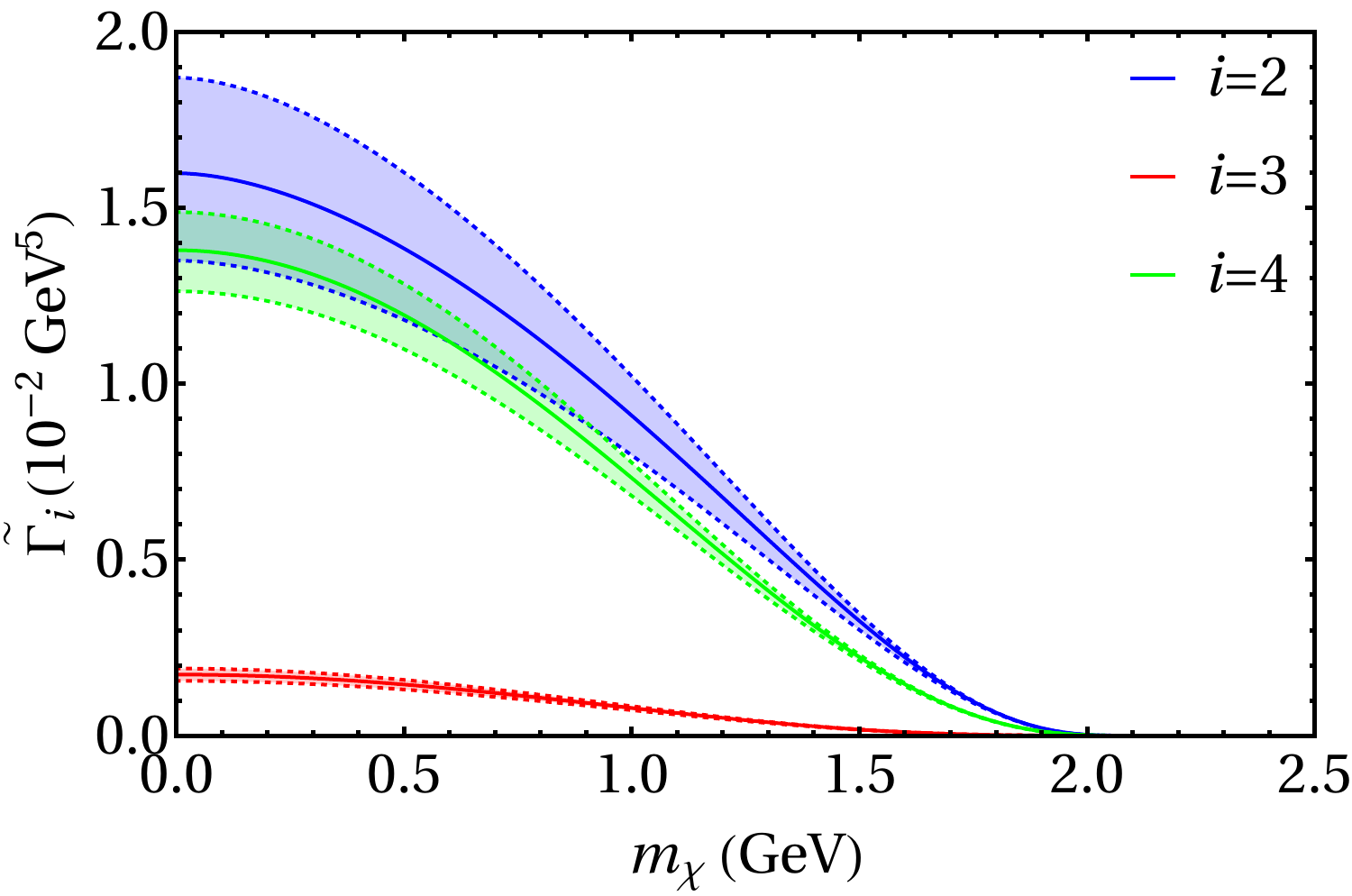}}\\
	\subfigure[$~B_c^-\to D^-$]{
		\includegraphics[width=0.43\textwidth]{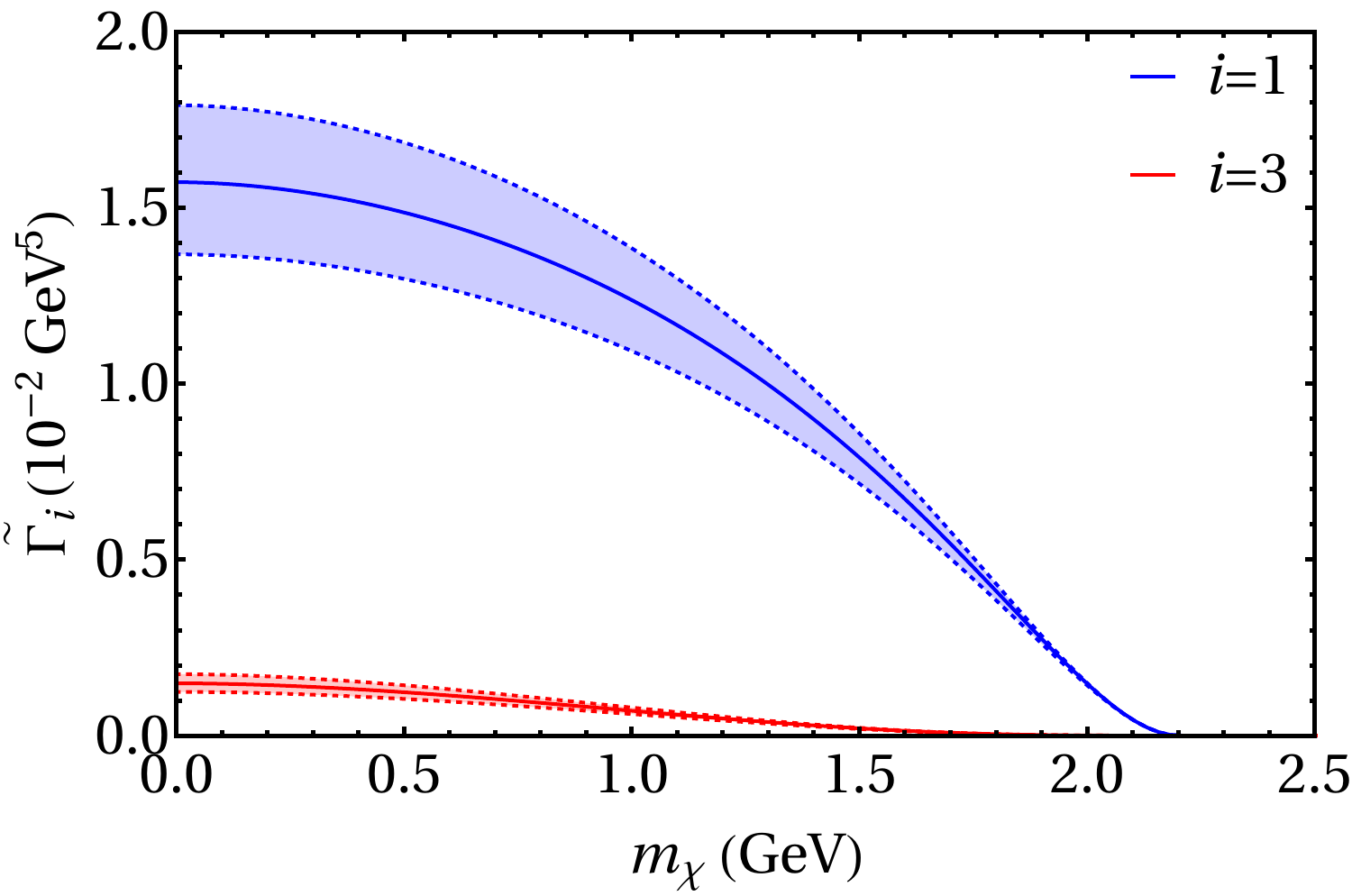}}
	\hspace{0.4cm}
	\subfigure[$~B_c^-\to D^{*-}$]{
		\includegraphics[width=0.43\textwidth]{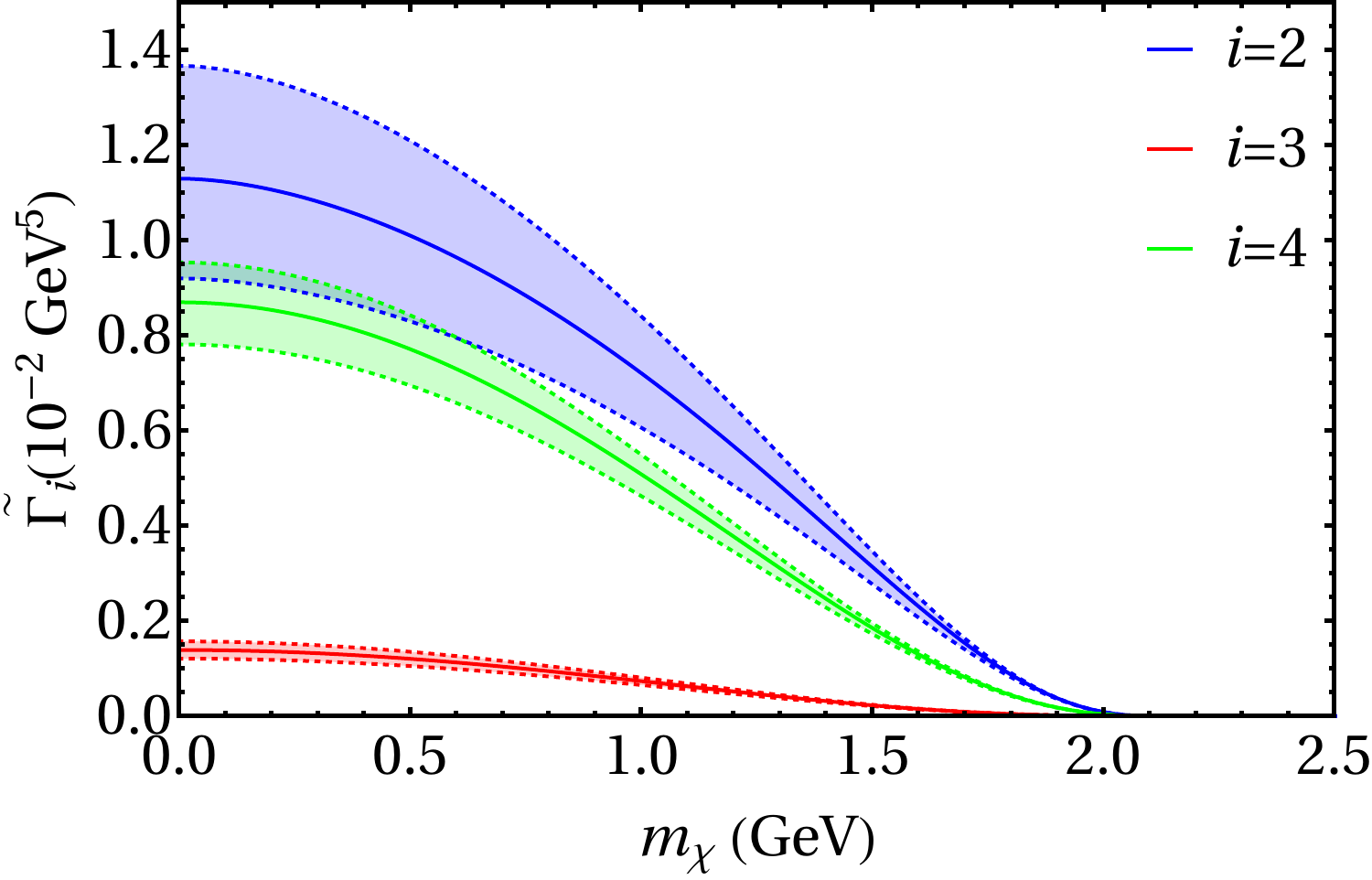}}\\
	\subfigure[$~B_c^-\to B^-$]{
		\includegraphics[width=0.42\textwidth]{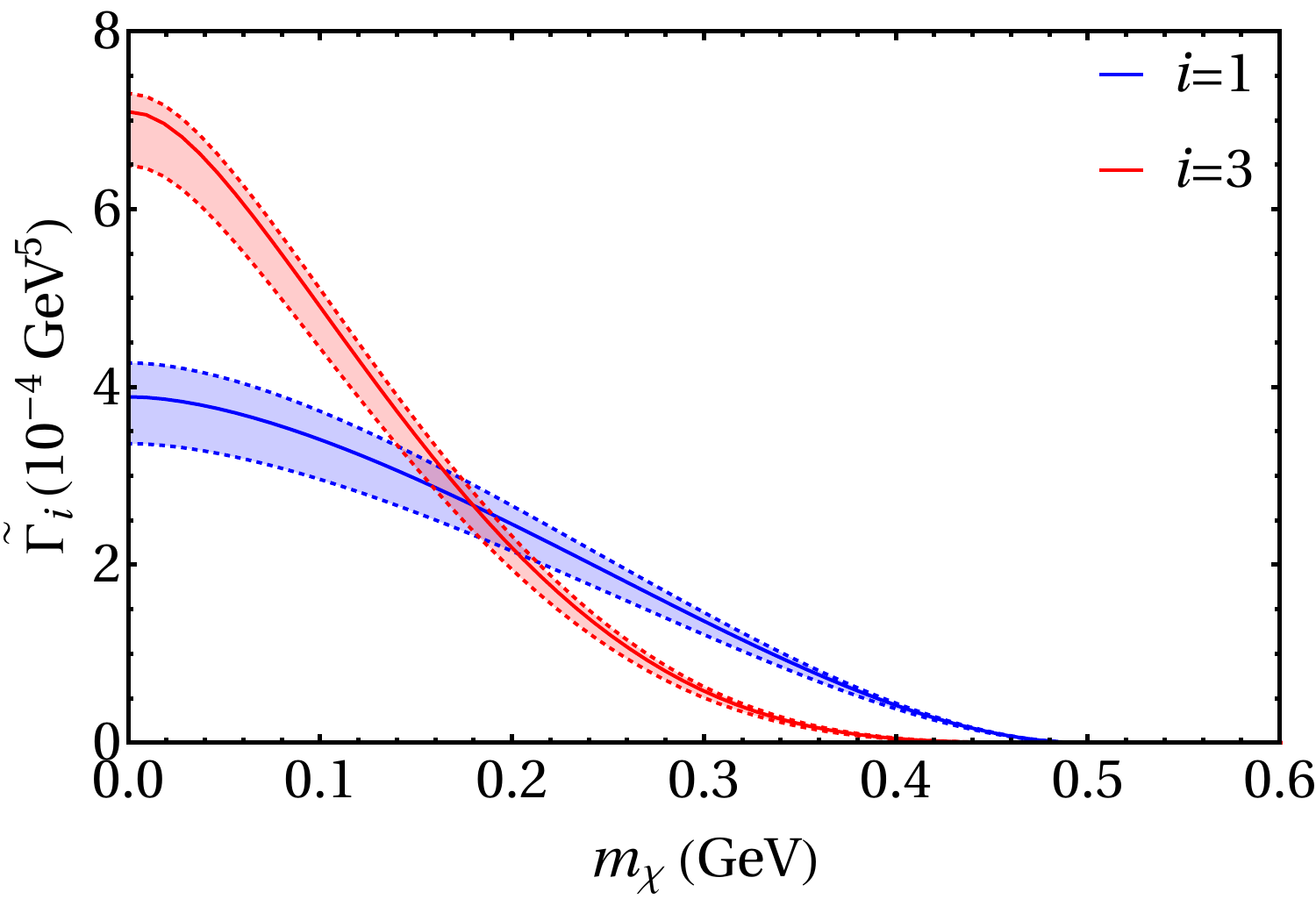}}
	\hspace{0.4cm}
	\subfigure[$~B_c^-\to B^{*-}$]{
		\includegraphics[width=0.43\textwidth]{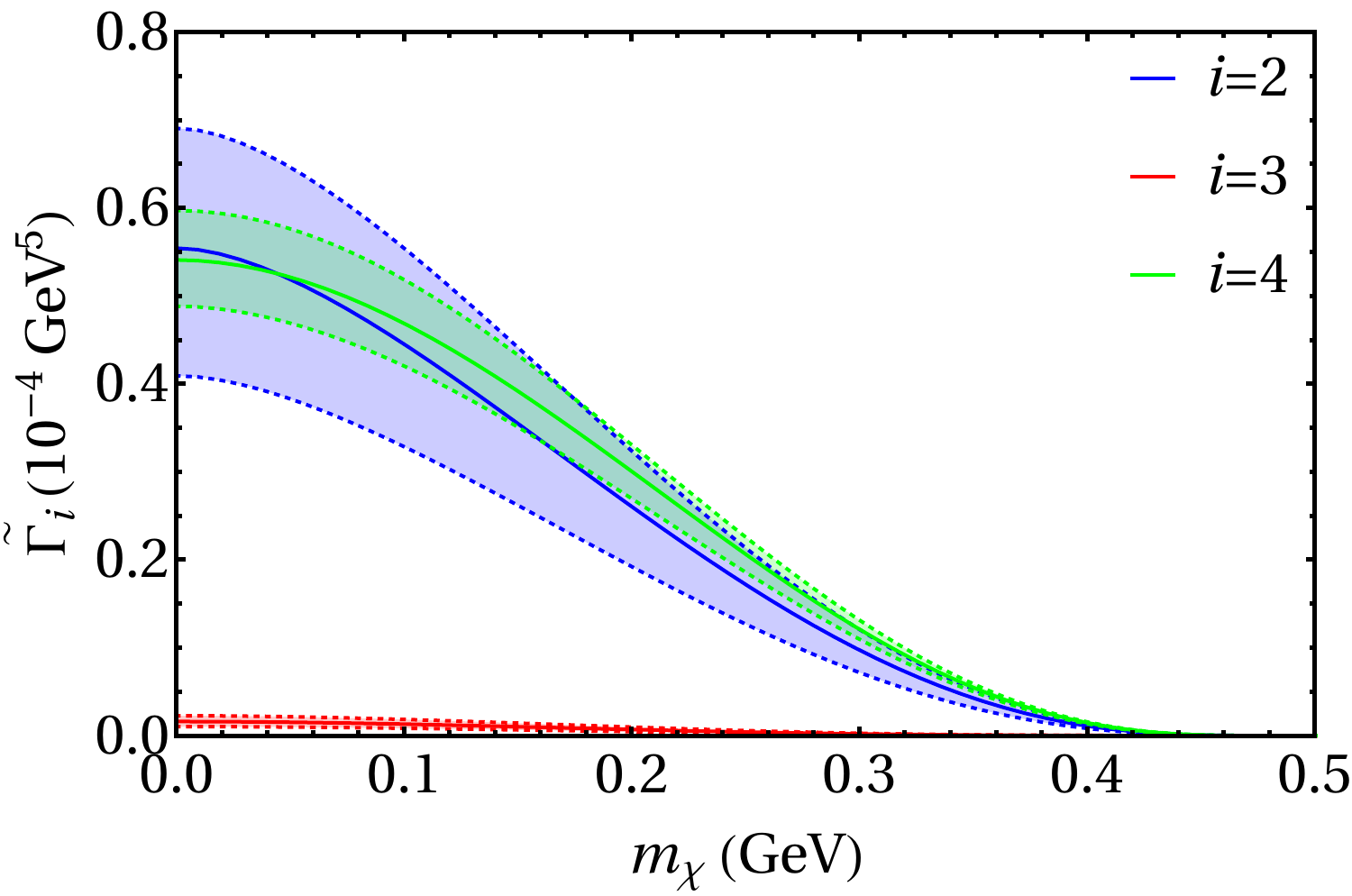}}
	\caption{The quantity $\widetilde\Gamma_i$ changes with $m_\chi$ in the $B_c^-\to P(V)\chi\chi^\dagger$ process. The shadows represent the errors estimated by varying the parameters in our model by $\pm5\%$.}
	\label{fig11}
\end{figure}

\begin{figure}[htb]
	\centering
	\subfigure[$B_c\to P$]{
		\label{fig12a}
		\includegraphics[width=0.43\textwidth]{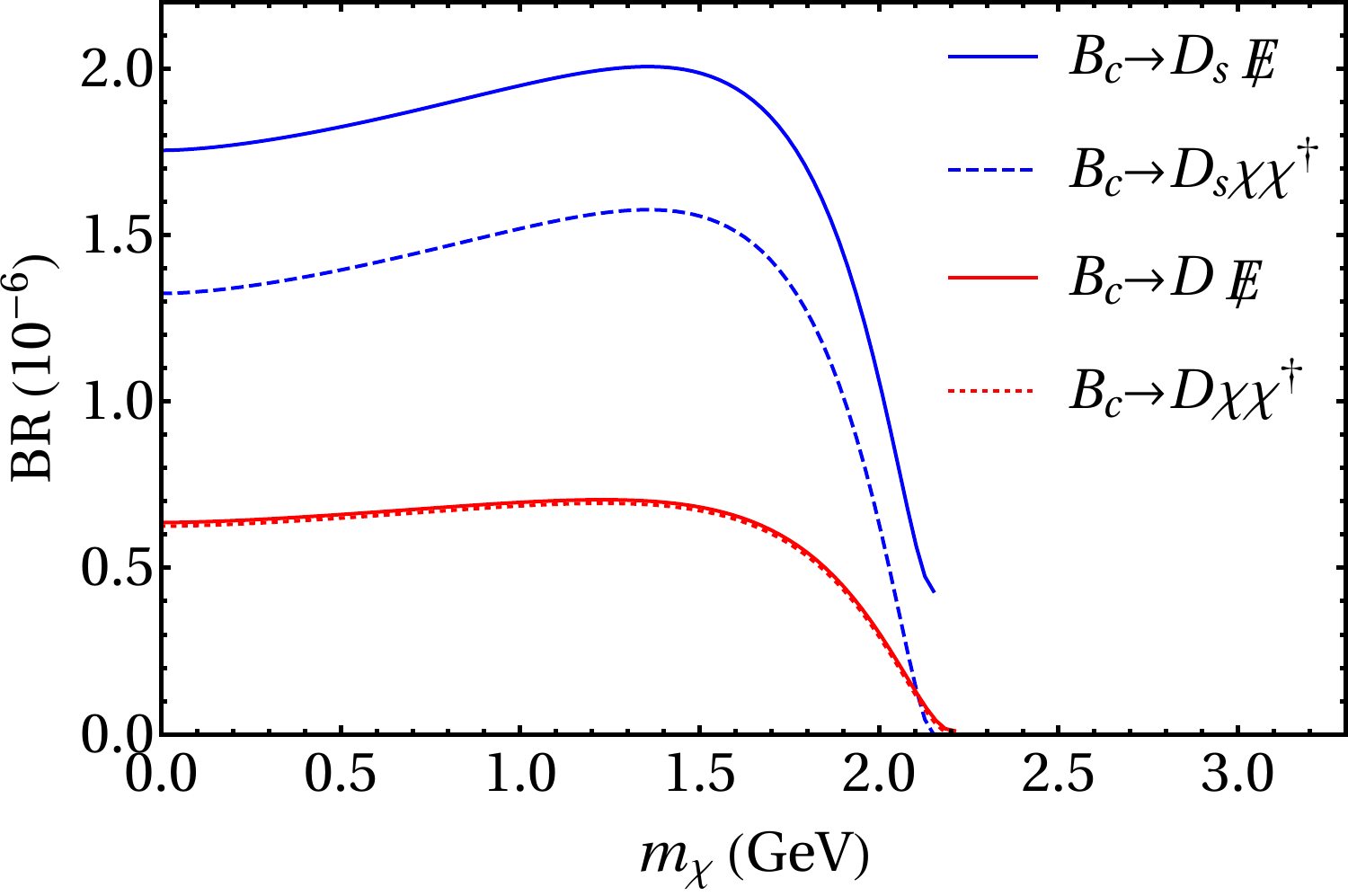}}
		\hspace{0.5cm}
	\subfigure[$B_c\to V$]{
		\label{fig12b}
		\includegraphics[width=0.42\textwidth]{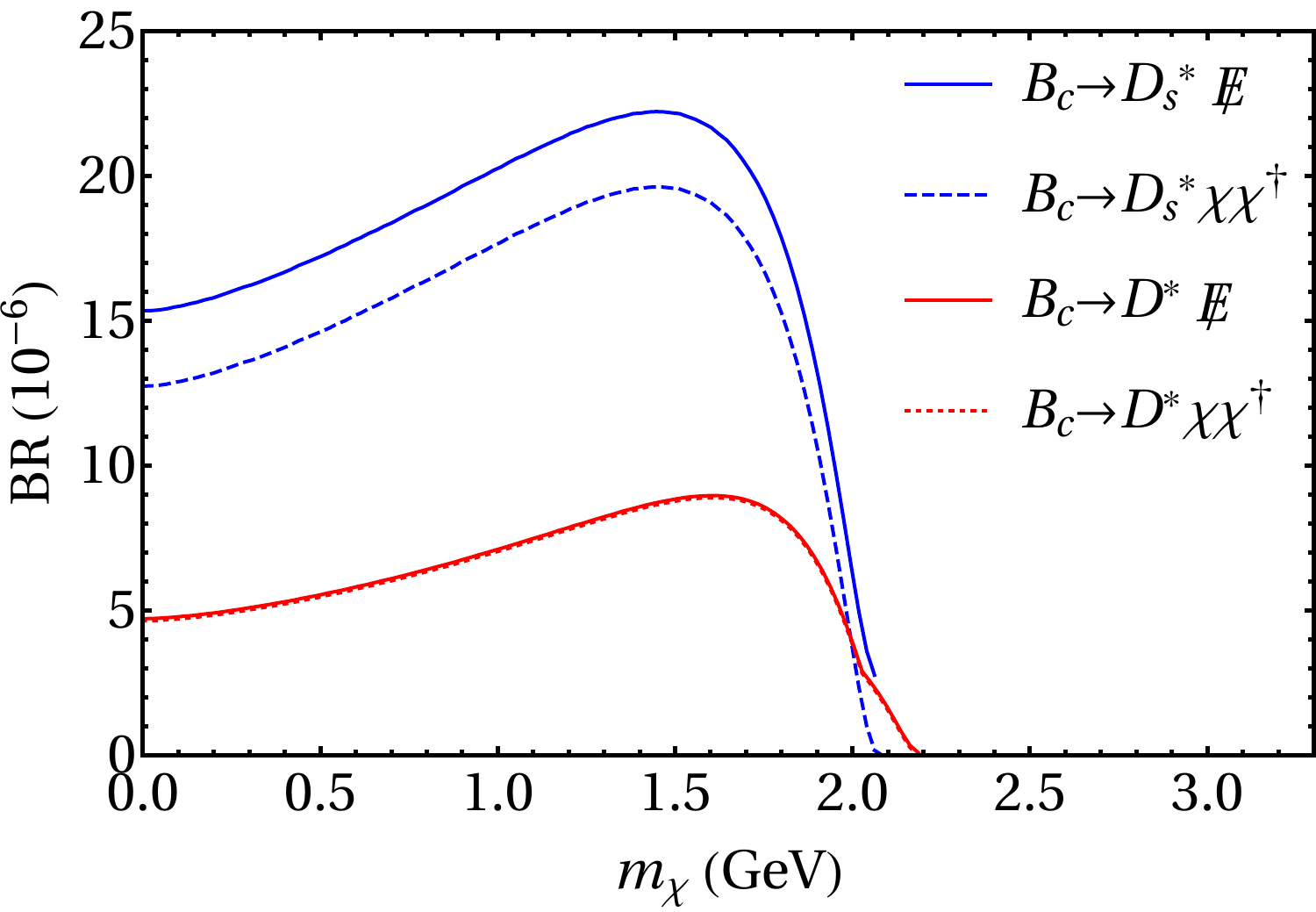}}
	\caption{Branching ratios of $B_c\to P(V)\chi\chi^\dagger$.}
	\label{fig12}
\end{figure}

One notices that, the Fig. \ref{fig12a} and Fig. \ref{fig10a} \ref{fig10b} are almost exactly the same though the calculation processes are very different. For the later, $\chi$ is a scalar particle, and only the scalar coupling operator takes effect. For the former, $\chi$ is a pseudoscalar, and two effective operators make contribution to the branching ratios. When we scan the parameter space in Fig. \ref{fig5}, which are right triangular regions, we find that if the right endpoint of the hypotenuse is taken, the branching ratio will achieve the maximum. This also means only the scalar operator should be considered. So when we calculate the upper limit of the branching ratios, the operator contributes to the decay modes of Fig. \ref{fig12a} is just the same as that contributes to Fig. \ref{fig10a} \ref{fig10b}, which makes their results are the same. But the condition in Fig. \ref{fig12b} is quite different. One can see the upper limits of the branching ratios are larger than those in Fig. \ref{fig10}. When we scan the parameter space in Fig. \ref{fig6}, we find the axial vector operator provides most of the contribution, which is different with the case in Fig. \ref{fig10c} \ref{fig10d}, where only the pseudoscalar operator takes effect.  For the $B_c^-\to D^{*-}\chi\chi^\dagger$ channel, there is a kink when $m_{\chi}$ is larger than $2$ GeV. This is because the corresponding operator which has the most important contribution turns to the vector from the axial vector.

\begin{figure}[htb]
	\centering
	\subfigure[$~B_c\to D_s$]{
		\label{fig13a}
		\includegraphics[width=0.43\textwidth]{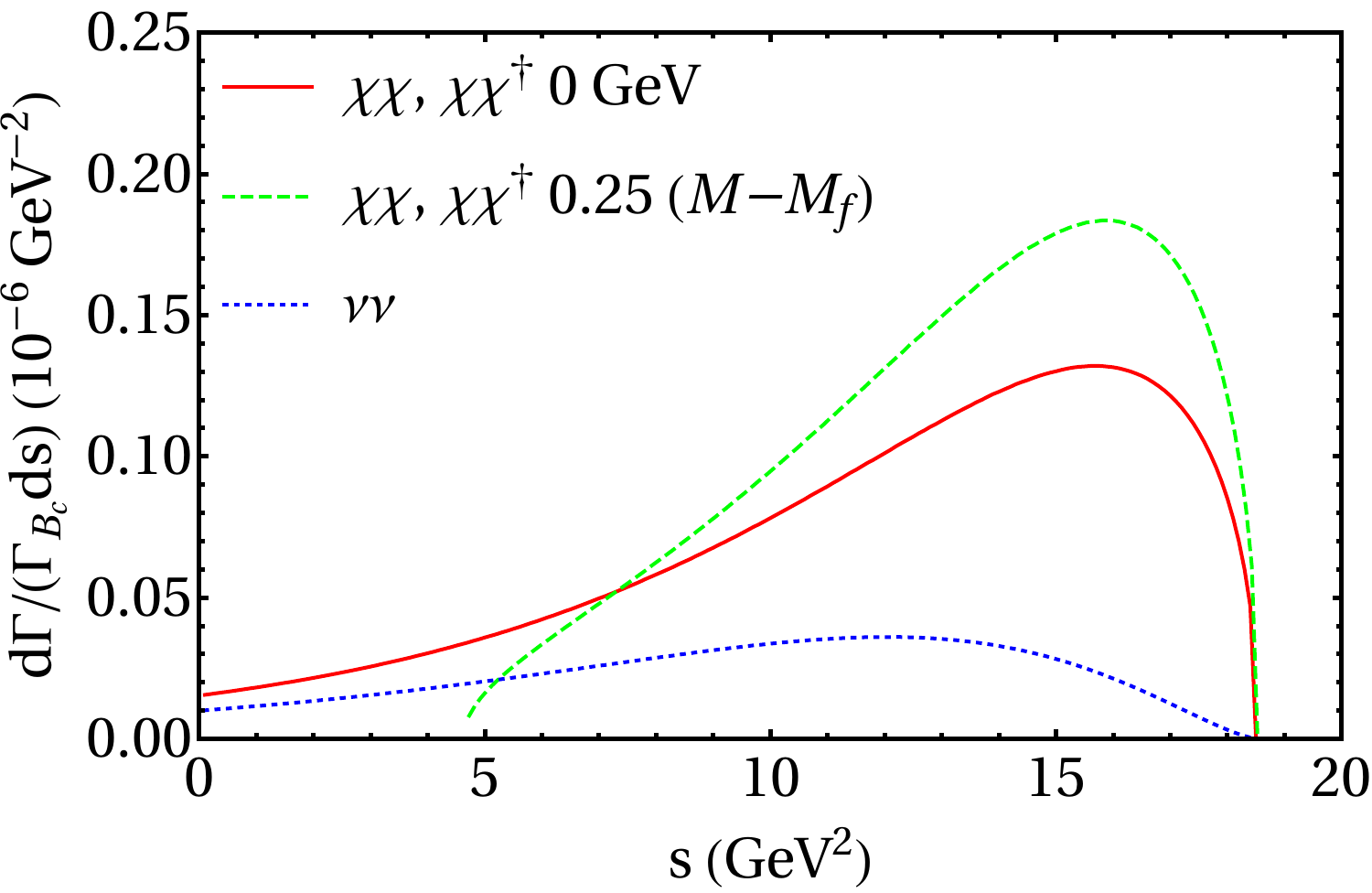}}
	\hspace{0.5cm}
	\subfigure[$~B_c\to D_s^\ast$]{
		\label{fig13b}
		\includegraphics[width=0.42\textwidth]{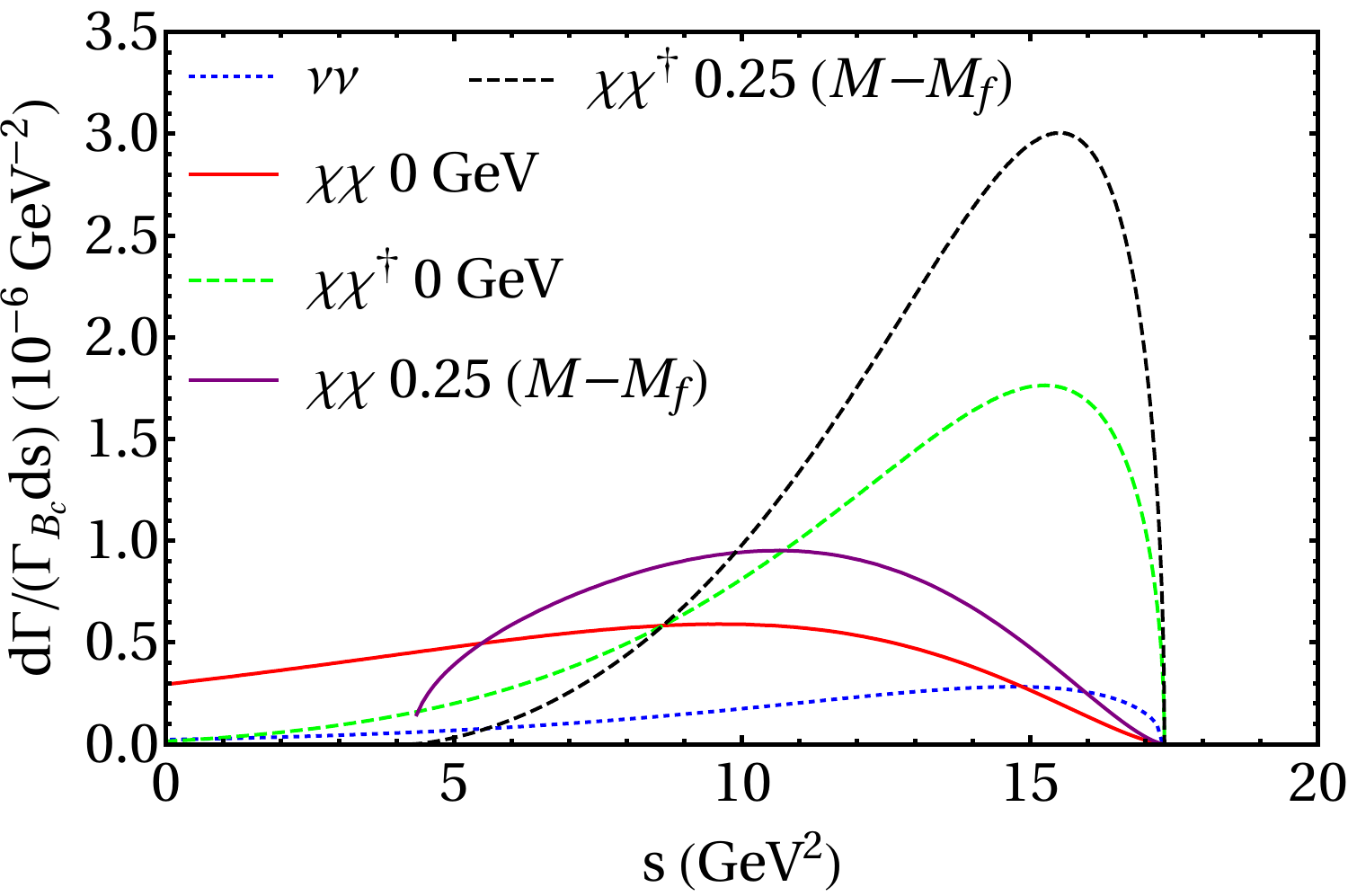}}\\
	\subfigure[$~B_c\to D$]{
		\label{fig13c}
		\includegraphics[width=0.43\textwidth]{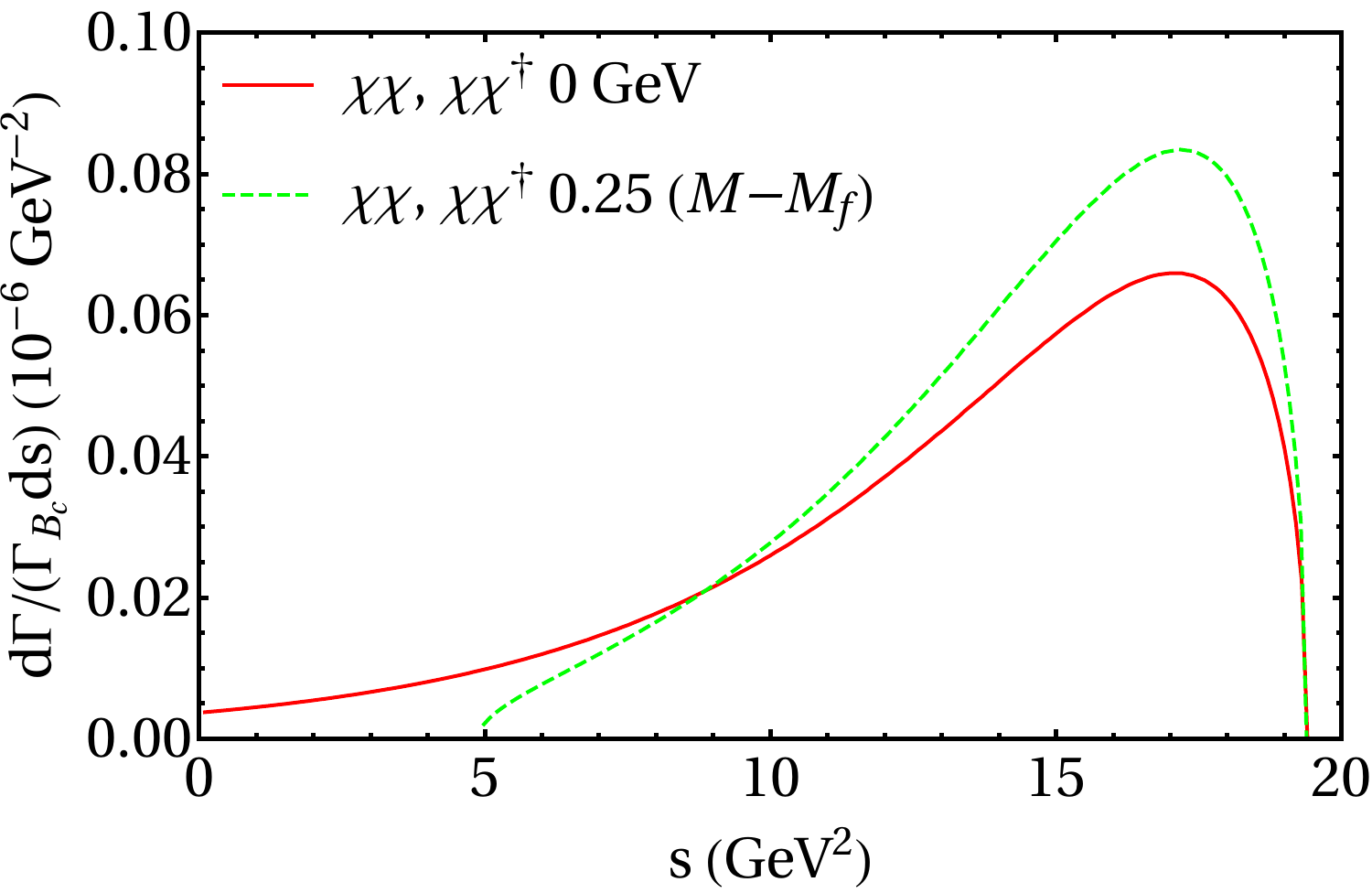}}
	\hspace{0.5cm}
	\subfigure[$~B_c\to D^\ast$]{
		\label{fig13d}
		\includegraphics[width=0.42\textwidth]{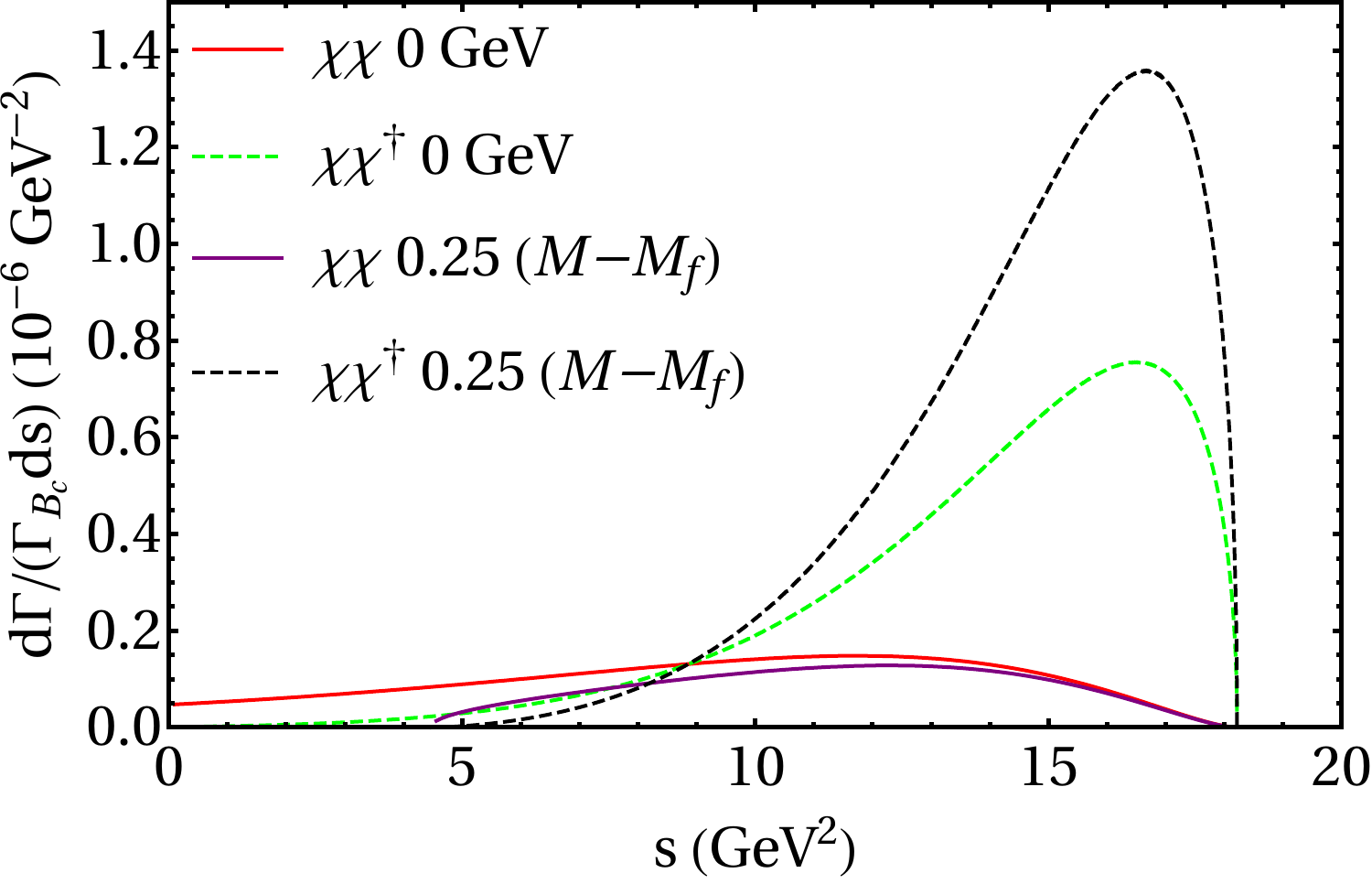}}\\
	\subfigure[$~B_c\to B^\ast$]{
			\label{fig13e}
		\includegraphics[width=0.43\textwidth]{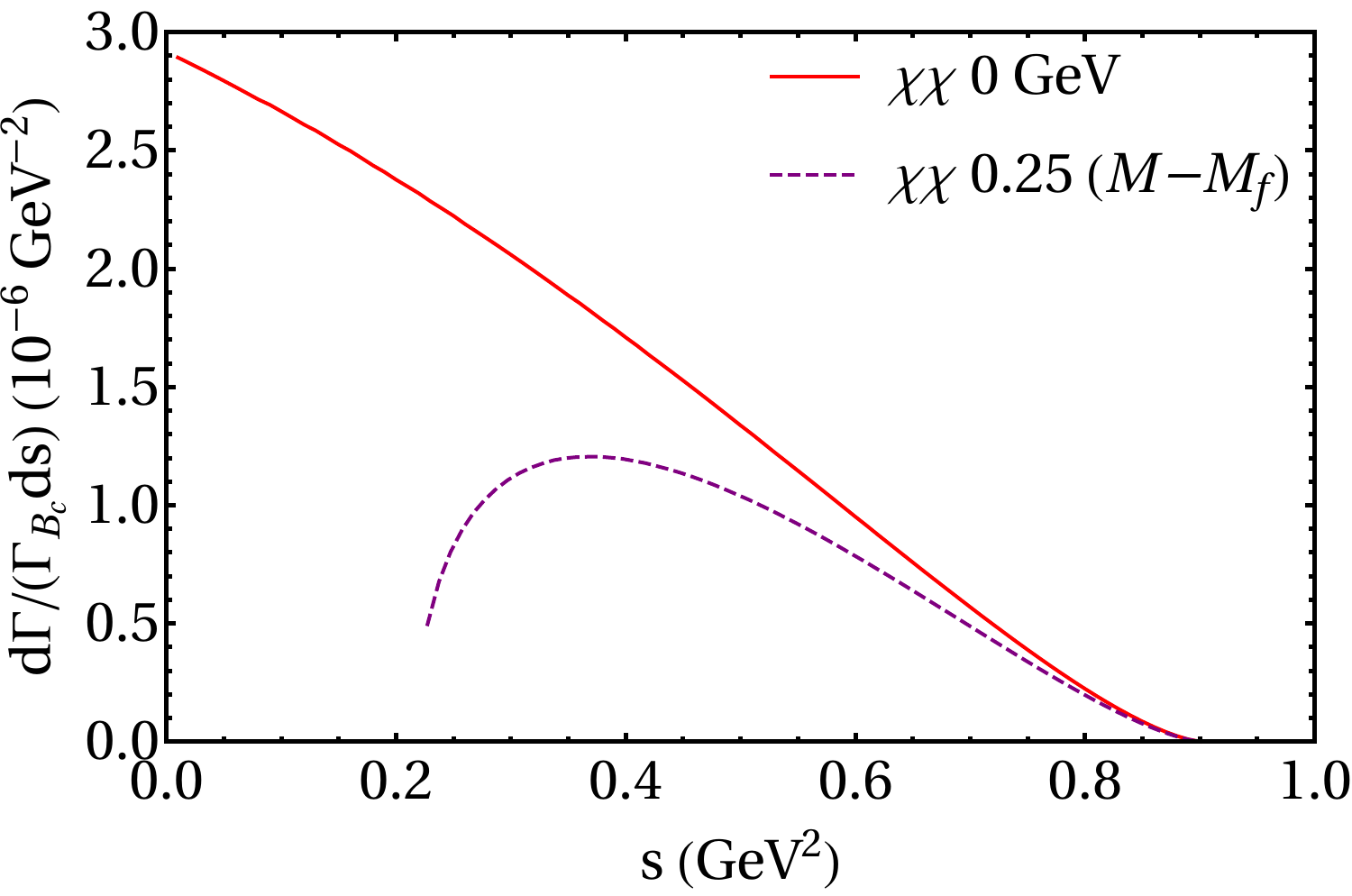}}
	\caption{Differential branching ratios of $B_c\to P(V)\nu\bar\nu$, $B_c\to P(V)\chi\chi$, and $B_c\to P(V)\chi\chi^\dagger$.}
	\label{fig13}
\end{figure}

In Fig. \ref{fig13} we present the differential distribution of the upper limits of the widths as a function of $s$. As examples, two cases with $m_\chi=0$ GeV and $0.25(M-M_f)$ are considered both for $\chi$ being a scalar or a pseudoscalar particle. In Fig. \ref{fig13a} and Fig. \ref{fig13c}, the lines for the decay modes with $\chi$ being a scalar or a pseudoscalar coincide. The reason for this is the same as that mentioned in the previous paragraph. The mass of $\chi$ determines the lower bound of $s$, namely, the left starting points of the curves. It is interesting to notice that for the $B_c\to D^{(\ast)}\chi\chi^\dagger$ and $B_c\to D_{s}^{(\ast)}\chi\chi^\dagger$ channels, the peaks of the distribution curves are always in the position $s=16$ GeV to $18$ GeV, which is almost independent of $m_\chi$.  The distribution curves for the $B_c\to B^\ast\chi\chi$ mode are a little bit different. When $m_\chi$ is very small, there is no peak. For comparison, we also plot the differential widths for the $B_c\to D_s^{(\ast)}\nu\bar\nu$ channels, which are smaller than those of invisible particles channels in most regions of $s$. For the channels $B_c\to D_s\nu\bar\nu$, $B_c\to D_s^\ast\nu\bar\nu$, and $B_c\to B^{(\ast)}\nu\bar\nu$, their decay widths are too small to be shown in Fig. \ref{fig13c}, Fig. \ref{fig13d}, and Fig. \ref{fig13e}, respectively.

\section{Conclusion}

We have studied the light invisible particles ($m_\chi$ is less than several GeV) through the rare decays of the $B_c$ meson. These particles can be the candidates of the light dark matter when the parameters taking specific values constrained by the $B$ meson decays and the relic density. Both the scalar and pseudoscalar cases are considered. Effective Lagrangians which contain the dimension-six operators are constructed to generate such processes. The effective coupling constants are constrained by the experimental results for the $B$ and $D$ decays with missing energy. Then the upper limits of the branching fractions of the $B_c\to P(V)\chi\chi$ and $B_c\to P(V)\chi\chi^\dagger$ channels are calculated. For the former, when the final meson is $D_s^{(\ast)}$, the largest value of the upper limits is of the order of $10^{-6}$; for the later, the largest value is of the order of $10^{-5}$ when the final meson is $D_s^\ast$. Although the results change with $m_\chi$, their orders of magnitude almost have no change if $m_\chi$ is not close to the threshold. Considering that the SM background is very small for some channels, we hope that the future experiments will find something new through such processes or set more stringent constraints for them.

\section{Acknowledgments}

This work was supported in part by the National Natural Science
Foundation of China (NSFC) under Grant No.~11405037, No.~11575048 and No.~11505039. We also thank the HPC Studio at Physics Department of Harbin Institute of Technology for access to computing resources through INSPUR-HPC@PHY.HIT.

\bibliography{reference}

\end{document}